\documentclass[aip,pop,reprint,amsmath,amssymb,superscriptaddress]{revtex4-1}

\usepackage{graphicx}
\usepackage{bm}
\renewcommand{\vec}[1]{\boldsymbol{#1}}

\def\grl{{Geophys. Res. Lett.} }
\def\jgr{{J. Geophys. Res.} }
\def\apj{{Astrophys. J.} }
\def\prl{{Phys. Rev. Lett.} }
\def\pop{{Phys. Plasmas} }
\def\ssr{{Space Science Reviews} }

\begin{document}

\title{Particle dynamics in the electron current layer in collisionless magnetic reconnection}

\author{Seiji Zenitani}
\affiliation{Division of Theoretical Astronomy, National Astronomical Observatory of Japan, 2-21-1 Osawa, Mitaka, Tokyo 181-8588, Japan. Electric mail: seiji.zenitani@nao.ac.jp.}
\author{Tsugunobu Nagai}
\affiliation{Tokyo Institute of Technology, Tokyo 152-8551, Japan}

\begin{abstract}
Particle dynamics in the electron current layer in collisionless magnetic reconnection
is investigated by using a particle-in-cell simulation.
Electron motion and velocity distribution functions are studied
by tracking self-consistent trajectories. 
New classes of electron orbits are discovered: 
figure-eight-shaped regular orbits inside the electron jet,
noncrossing regular orbits on the jet flanks,
noncrossing Speiser orbits, and
nongyrotropic electrons in the downstream of the jet termination region. 
Properties of a super-Alfv\'{e}nic outflow jet are attributed to
an ensemble of electrons traveling through Speiser orbits. 
Noncrossing orbits are mediated by the polarization electric field
near the electron current layer.
The noncrossing electrons are found to be non-negligible in number density.
The impact of these new orbits to electron mixing,
spatial distribution of energetic electrons, and
observational signatures, is presented.
\end{abstract}

\maketitle

\section{Introduction}

Collisionless magnetic reconnection is
a basic plasma process for the abrupt release of magnetic energy.
Understanding the process is crucial to discuss
planetary magnetospheres, solar corona, the solar wind,
laboratory plasmas, and astrophysical plasma environments.
Collisionless reconnection is a highly nonlinear, complex process,
in which the electromagnetic field and the plasma particle motion
interact with each other. 
The reconnection mechanism has not yet been fully understood, but
numerical simulations provide a way to investigate the underlying physics.

Since early research with particle-in-cell (PIC) simulations,
it has been recognized that
a small-scale electron-physics layer is embedded
inside a broader ion-physics layer
in collisionless magnetic reconnection.
For example, \citet{prit01a} presented
a narrowly collimated electron jet inside a broader ion outflow.
This picture was further extended by successive PIC simulations,
\citep{dau06,keizo06,kari07,shay07,drake08,klimas08}
which were large enough to separate electron-scale structure from the ion-scale structure.
\citet{kari07} and \citet{shay07} demonstrated that
the electron-physics layer
evolves into the inner core region and the fast elongated jet. 
Although these results raised a question
concerning long-term behavior of magnetic reconnection,\citep{klimas08}
now many scientists agree that
the inner core region controls the reconnection rate.
The entire layer is often called
the electron current layer (ECL) or the electron diffusion layer.
Hereafter we call it the electron current layer (ECL).

The inner core is called the dissipation region (DR)
or the electron diffusion region (EDR).
This is the site of dissipation physics,
arising from complex electron motions
(See \citet{hesse11} for a review).
As of today, the DR is ambiguously defined, and
there are many different opinions on its rigorous definition.\citep{goldman16}
Promising signatures to identify the DR are
enhanced energy dissipation,\citep{zeni11c}
electron nongyrotropic behavior,\citep{scudder08,aunai13c,swisdak16}
electron phase-space hole along the inflow direction,\citep{hori08,chen11}
and characteristic velocity distribution functions (VDFs).\citep{ng11,bessho14}

The jet is popularly referred to as
the super-Alfv\'{e}nic electron jet,
because its bulk speed exceeds the Alfv\'{e}n speed in the inflow region.
This fast jet has characteristic features such as
violation of the electron ideal condition,\citep{prit01a,kari07,shay07}
diamagnetic-type momentum balance,\citep{hesse08}
an electron pressure anisotropy,\citep{le10a}
electron nongyrotropy,\citep{scudder08,aunai13c,swisdak16}
bipolar polarization electric fields $E_z$
(the so-called Hall electric fields),
and highly structured electron VDFs.\citep{aunai13c,bessho14,shuster15}
These issues are usually discussed separately.
The number of attempts to comprehensively explain
these signatures has been limited.

There have been many observational studies on the ECL
during magnetic reconnection in the Earth's magnetosphere.
\citet{chen08,chen09} studied a magnetotail reconnection event with Cluster spacecraft.
With the help of a map of electron VDFs by PIC simulation,
they identified that the satellite crossed
an electron-scale thin current layer near the X-line.
\citet{nagai11} reported an informative reconnection event
in the magnetotail with the Geotail spacecraft. 
They detected both bi-directional electron flows that outrun ion flows
and an energy dissipation site around the X-line.\citep{zeni12}
\citet{nagai13b} studied another magnetotail reconnection event
with ion-electron decoupling.
They reported additional signatures
in the ion-electron decoupling region,
such as the energetic electron fluxes. 
\citet{oka16} reported a DR-crossing in the magnetotail with THEMIS.
They observed nongyrotropic electron VDFs and perpendicular heating inside the ECL.
They also showed that the ECL is a site of electron energetization.
In addition,
there have been several observations of super-Alfv\'{e}nic electron jets
inside the reconnection outflow exhaust
in the magnetosheath,\citep{phan07}
in the magnetotail,\citep{zhoum14}
and
in the solar wind.\citep{xux15}
These results support the standard picture of the ECL,
the central DR and extended electron jets.

In the above observations,
time and spatial resolutions were rather limited
to discuss electron physics in great detail.
In order to probe electron-scale structures
in near-Earth reconnection sites
at ultra-high resolutions,
NASA recently launched the Magnetospheric MultiScale (MMS) spacecraft
in 2015.\citep{burch16}
MMS is planned to move to the Earth's magnetotail in 2017,
where magnetic reconnection often occurs
in an anti-parallel configuration. 
It is expected that
MMS will encounter ${\approx}10$ reconnection events.\citep{kevin14}
Once it encounters reconnection events,
MMS is expected to resolve the aforementioned structures.

Since spacecraft observe VDFs,
it is important to understand the electron motion behind the VDFs
in PIC simulations.
\citet{hoshino01b} was one of the first to discuss
electron VDF and associated particle motion.
They examined VDFs near the magnetic island in the downstream region. 
\citet{egedal05,egedal08} found that
electron VDFs are elongated in the field-aligned direction
in the inflow region.
As already introduced, \citet{chen08,chen09} studied
spatial distribution of electron VDFs by PIC simulation
to interpret satellite data. 

In order to prepare for the MMS observation,
a growing number of works are devoted to electron VDFs in PIC simulations.
\citet{ng11} visualized
the complex structure of the electron VDF in the DR. 
They reconstruct the VDF at high resolution,
by back-tracing particle orbits in the PIC field and
by using a Liouville's theorem.
The VDF contains discrete striations in a triangular-shaped envelope.
Shuster et al.\citep{shuster14,shuster15} presented
that electron VDFs contain various discrete components
in the outflow exhaust.  They further examined
the structure of the electron VDF over the ECL,
with help from test-particle simulations.
\citet{bessho14} examined electron VDFs in the ECL.
In the DR, they gave semi-analytic expressions to
the fine structure of the VDF. 
In the downstream, many arcs were found in the VDFs
and they were attributed to a gradual remagnetization of electrons.
\citet{frankcheng15} inferred
particle dynamics in driven reconnection
from the spatial distribution of VDFs.
They claimed that
the field-aligned electron population in the inflow region is
injected into the super-Alfv\'{e}nic electron jet.
Most recently, \citet{wang16a} studied
the electron heating mechanism in the exhaust region in detail.
With help from test particle simulations,
they found that
parallel heating by the curvature drift acceleration and
perpendicular heating by the gradient-B drift acceleration
account for a highly structured VDF near the magnetic flux pile-up region.

These VDFs are ensembles of electrons,
following various complex trajectories
in the reconnection system.
In addition to a gyration and a parallel motion,
several classes of electron particle motions
are reported in the previous literature,
such as
a Speiser motion around the DR\citep{speiser65,ng11,bessho14}
and a field-aligned bounce motion in the inflow region.\citep{egedal05,egedal08}
They are proven to be building blocks of the VDFs.
However, it is not clear
whether these particle motions and/or their combinations
can explain everything about electron VDFs. 
In fact, as reviewed in this section,
the ECL structure is found to be much more complicated than
previously expected.
It is possible that
some electrons travel through new orbits near the reconnection site
and that
they have an impact on the reconnection dynamics and observational signatures.
In order to better interpret the electron VDFs
to deeper discuss kinetic reconnection physics,
it is important to understand electrons particle orbits and dynamics
in a modern reconnection simulation.

The purpose of this paper is
to comprehensively discuss
electron fluid properties, VDFs,
self-consistent trajectories, and
relevant particle dynamics
around the ECL in magnetic reconnection. 
The paper is organized as follows.
First, we briefly review nongyrotropic particle motions
in a curved magnetic field in Section \ref{sec:theory}.
Next, we describe
the numerical setup of a 2D PIC simulation in Section \ref{sec:setup}.
The simulation results are presented in several ways.
Section \ref{sec:fluid} presents
macroscopic fluid quantities.
In Section \ref{sec:kinetic},
we present electron kinetic signatures such as VDFs and phase-space diagrams.
In Section \ref{sec:traj},
we will discuss self-consistent electron trajectories in detail.
In Section \ref{sec:comp},
we utilize the trajectory datasets
to further examine VDFs and
spatial distribution of electrons.
The relevance to the satellite observation is briefly
addressed in Section \ref{sec:ET}.
Section \ref{sec:discussion}
contains discussions and summary.

\section{Electron motion in a curved field reversal}
\label{sec:theory}

We outline basic properties of
particle (electron) trajectories
in a highly bent magnetic field.\citep{chen86,BZ89}
Here we consider a simple parabolic field,
\begin{eqnarray}
\label{eq:B}
\vec{B}=B_0(z/L)\vec{e}_x+B_n\vec{e}_z, ~~\vec{E}=0,
\end{eqnarray}
where 
$B_0$ is the reference magnetic field,
$L$ is the length scale of the current sheet, and
$B_n$ is the normal magnetic field.
This system resembles the outflow region
in magnetic reconnection at the lowest order.
The equation of motion,
$m_e({d\vec{v}_e}/{dt}) = -e(\vec{v}_e\times\vec{B})$,
can be rewritten
\setcounter{equation}{2}
\begin{align}
~~~~~~~~~~~~~~~
\left\{
\begin{aligned}
\ddot{x} &=& &-\Omega_n \dot{y}                             & ~~~~~~~~~~~~{\textrm (2a)} \\
\ddot{y} &=& -\omega_b^2(\dot{z}/|v_e|)z &+\Omega_n \dot{x} & {\textrm (2b)} \\
\ddot{z} &=&~~\omega_b^2(\dot{y}/|v_e|)z &                  & {\textrm (2c)}
\end{aligned}
\right.
\notag
\end{align}
where $\Omega_n=eB_n/m_e$ is the gyrofrequency about $B_n$ and
$\omega_b=\sqrt{eB_0|v_e|/m_eL}$ is a characteristic frequency. 
The electron motion is characterized by
the ratio of the two frequencies,\citep{BZ89} 
\begin{align}
\label{eq:kappa}
\kappa
\equiv
\frac{\Omega_n}{\omega_b}
=
\sqrt{\frac{R_{\rm c,min}}{r_{\rm L,max}}}
=
\Big|\frac{B_n}{B_0}\Big|\sqrt{\frac{L}{r_{\rm L,max}}}
,
\end{align}
where
$R_{\rm c,min}$ is the minimum curvature radius of magnetic field line
and
$r_{\rm L,max} = m_e|v_e| / eB_n$ is the electron's maximum gyroradius.
This parameter is known as the curvature parameter.
The curvature radius solely depends on the field-line geometry,
while the gyroradius $r_{\rm L,max}$ depends on
the electron velocity $|v_e|$ or the electron energy $\mathcal{E}\equiv\frac{1}{2}m_ev_e^2$.
Note that both $|v_e|$ and $\mathcal{E}$ are constant
in this system, because $\vec{E}=0$. 

When $\kappa \gg 1$, the gyration about $B_n$ dominates.
If the parameter falls below the unity, $\kappa \lesssim 1$,
the electron motion becomes nongyrotropic. 
In particular, the motion becomes highly chaotic
for $\kappa \sim 1$ ($\Omega_n \sim \omega_b$),
because the two oscillations of different kinds interfere with each other.
Several characteristic orbits appear for $\kappa \ll 1$,
as will be shown in the next paragraphs.

\begin{figure}[btp]
\centering
\includegraphics[width={0.48\textwidth},clip]{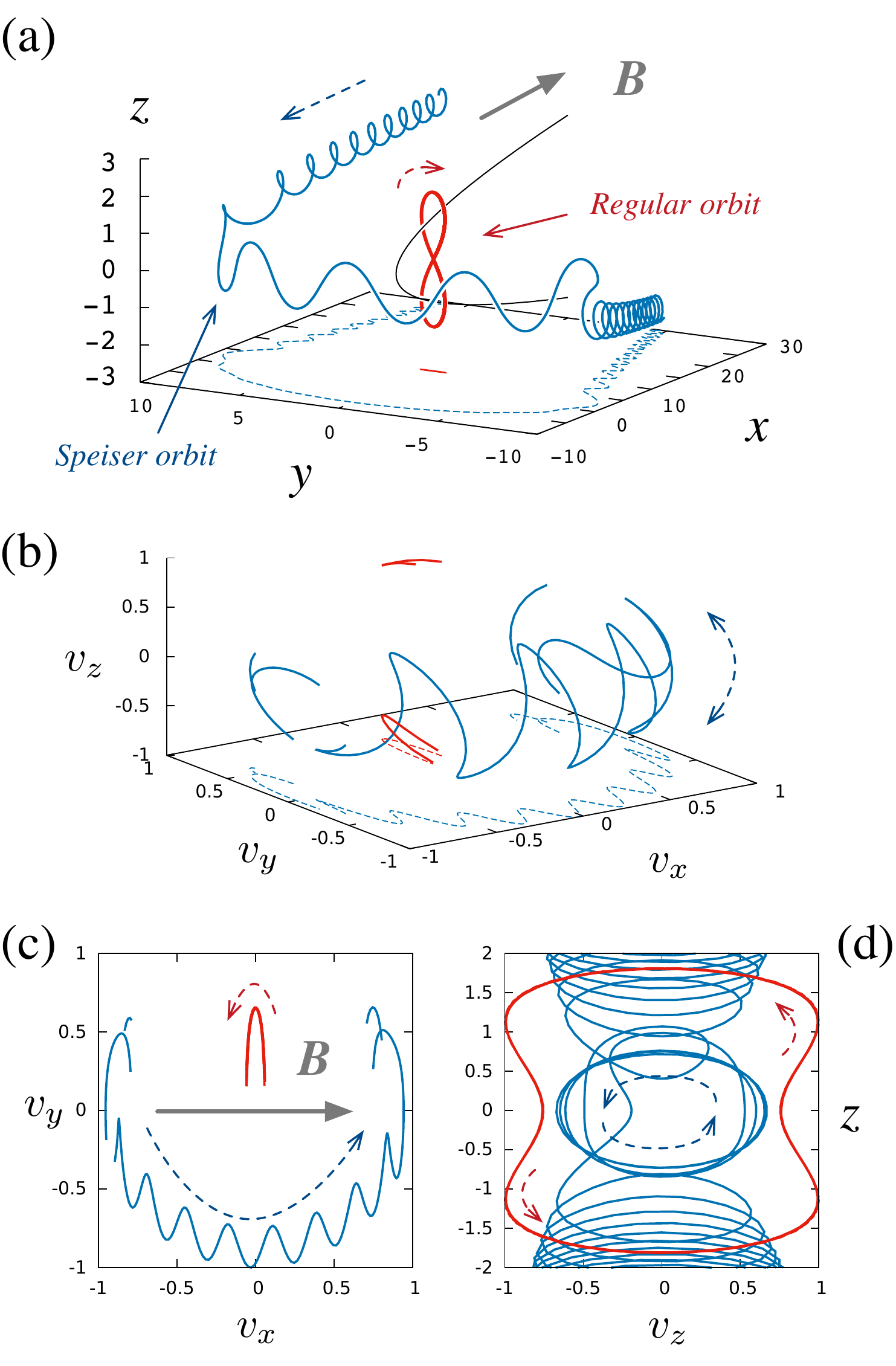}
\caption{(Color online)
(a) Nongyrotropic electron orbits in a curved field reversal for $\kappa=0.1$.
A Speiser orbit (blue) and a regular orbit (red).
(b) 3D velocity-space orbits of the electrons near the midplane $|z|<1$.
(c) 2D velocity-space orbits in the $v_x$--$v_y$ plane.
(d) Phase-space diagram for the Speiser and regular orbits in $v_z$--$z$.
}
\label{fig:theory}
\end{figure}

Figure \ref{fig:theory} demonstrates
typical electron orbits for $\kappa=0.1$
in the 3D space (Fig.~\ref{fig:theory}a), the velocity space (Figs.~\ref{fig:theory}b and \ref{fig:theory}c),
and the phase-space ($v_z$--$z$; Fig.~\ref{fig:theory}d).
Parameters are chosen to be $|R_c|=1$ and $|v_e|=1$. 
The blue orbit demonstrates
a well-known Speiser orbit.\citep{speiser65}
After entering the midplane from the upper left,
it slowly turns its direction from $-x$ to $+x$
due to the gyration about $B_n$.
Then it exits in the $+x$ direction. 
Near the midplane ($z\sim 0$),
the electron mainly travels in $-y$,
while bouncing in $z$ at the frequency of $\omega_b$
($\dot{y}\approx -|v_e|$ in Eq.~(2c)).
This is the so-called meandering motion.
In the velocity space,
the $B_n$-gyration near the midplane
corresponds to a half circle in $v_x$--$v_y$,
as can be seen in Figure \ref{fig:theory}c.
The electron velocity rotates anti-clockwise from $-v_x$ to $+v_x$.
The fast $z$-bounce motion is evident in ${\pm}v_z$,
as indicated by the arrow in Figure \ref{fig:theory}b.
Since the meandering motion consists of two opposite gyrations,
the electron moves back-and-forth along arcs in ${\pm}v_z$.
It finally exhibits a zigzag pattern in the 3D velocity space.
The $z$-bounce motion also corresponds to
the rotation around the center
in the phase-space (Fig.~\ref{fig:theory}d).

The red orbit belongs to another kind of nongyrotropic orbit.
It is called a regular orbit or an integrable orbit.\citep{chen86}
When the $z$-motion fully resonates with the gyration about $B_n$,
the electron travels through a figure-eight-shaped orbit,
hitting a fixed point at the midplane
(See also the left panel of Fig.~4 in Ref.~\onlinecite{chen86}).
In the vicinity of the figure-eight-shaped orbit,
electrons keep bouncing in $z$ and do not escape away from the midplane.
It was further shown that
the electrons are trapped on the surface of a ring-type torus, but
the reader is referred to \citet{BZ89} paper for detail.
Because of the $z$-bounce motion,
we usually see closed circuits in the phase space (Fig.~\ref{fig:theory}d).
Note that the regular-orbit electrons travel in $+v_y$
near the midplane ($z \sim 0$).
They appear in the $+v_y$ side in the velocity space
near the midplane (Fig.~\ref{fig:theory}c).

\section{Simulation}
\label{sec:setup}

We use a partially implicit PIC code\citep{hesse99}
to study our reconnection problem.
The length, time, and velocity are normalized by
the ion inertial length $d_i=c/\omega_{pi}$,
the ion cyclotron time $\Omega_{ci}^{-1}=m_i/(eB_0)$, and
the ion Alfv\'{e}n speed $c_{Ai}=B_0/(\mu_0 m_i n_0)^{1/2}$, respectively.
Here, $\omega_{pi}=( e^2n_0/\varepsilon_0 m_i)^{1/2}$ is the ion plasma frequency,
and $n_0$ is the reference density.
We employ a Harris-like configuration,
$\vec{B}(z)=B_0 \tanh(z/L) \vec{\hat{x}}$ and
$n(z) = n_{0} \cosh^{-2}(z/L) + n_b$,
where the half thickness is set to $L=0.5 d_i$ and
$n_b = 0.2 n_0$ is the background density.
The ion-electron temperature ratio is $T_i/T_e=5$.
The mass ratio is $m_i/m_e=100$.
The ratio of the electron plasma frequency to the electron cyclotron frequency
is $\omega_{pe}/\Omega_{ce}=4$.
Our domain size is $x,z \in [0,76.8]\times[-19.2, 19.2]$.
It is resolved by $2400 \times 1600$ grid cells.
Periodic ($x$) and reflecting wall ($z$) boundaries are employed.
$1.7{\times}10^9$ particles are used.
Reconnection is triggered by a small flux perturbation,
$\delta A_y = - 2L B_1 \exp[-(x^2+z^2)/(2L)^2]$,
where $B_1=0.1 B_0$ is the typical amplitude of the perturbed fields.
The initial electric current is configured accordingly.

This run was analyzed in our previous paper
on ion VDFs and ion particle dynamics.\citep{zeni13}
The parameters and system evolution are similar to
those in run 1A in Ref.~\onlinecite{zeni11d}
on the electron-scale structure.
Several aspects of this reconnection system were
presented in these papers. 
We explore new aspects of electron VDFs and particle dynamics
in this paper.

\begin{figure*}[tbp]
\centering
\includegraphics[width={\textwidth},clip]{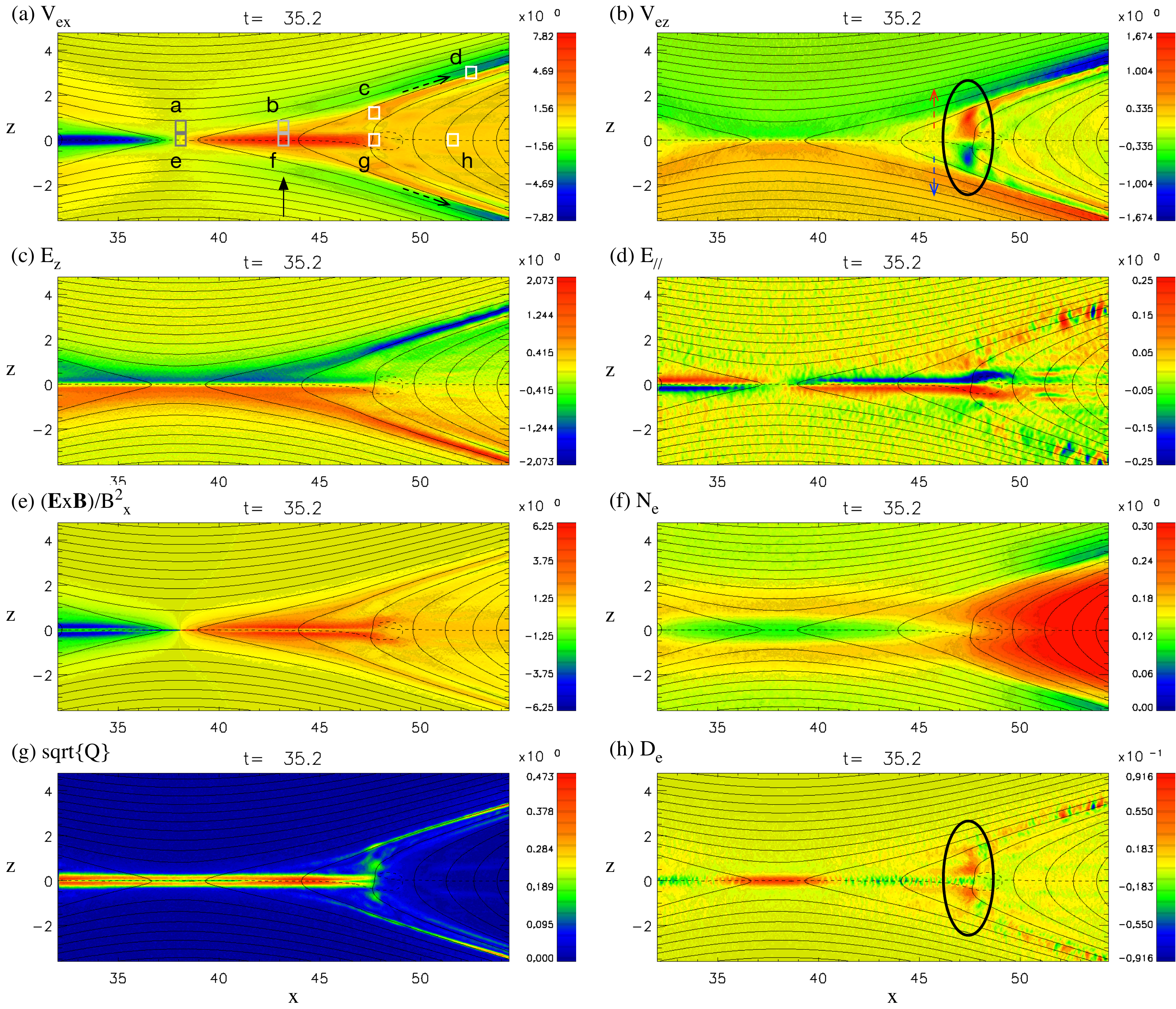}
\caption{(Color online)
Results of the main run, averaged over $t=35$--$35.25$.
The contour lines are in-plane magnetic field lines and
the dashed line indicates the field reversal, $B_x=0$.
(a) The electron outflow speed $V_{ex}$ in a unit of $c_{Ai}$,
(b) the vertical electron flow $V_{ez}$,
(c) the vertical electric field $E_z$ in a unit of $c_{Ai}B_0$,
(d) the parallel electric field $E_{\parallel}$,
(e) the {\bf E}$\times${\bf B} outflow speed $w_x$,
(f) the electron density $n_{e}$,
(g) the nongyrotropy measure $\sqrt{Q}$, and
(h) the nonideal energy dissipation $\mathcal{D}_e$.
}
\label{fig:snapshot}
\end{figure*}

\section{Fluid quantities}
\label{sec:fluid}

The reconnection occurs at the center of the simulation domain.
The magnetic flux transfer rate across the X-line grows in time until $t \approx 14.5$.
If normalized by quantities at $3 d_i$ upstream of the X-line,
the reconnection rate reaches $0.14$, gradually decreases to $0.11$ at $t \approx 25$,
and then remains constant after that.
An electron jet as well as other electron-scale structures grows in time.
They are well developed at the time of our interest, $t=35$.
We study this time step,
because we studied other aspects in our previous studies,\citep{zeni11d,zeni13}
and
because it is early enough to avoid major effects from the periodic boundary in $x$.
In fact, the electron jet continues to evolve until $t \approx 44$,\citep{zeni11d}
but minor boundary effects appear in particle signatures at $t \approx 35$,
as will be shown later.

Figure \ref{fig:snapshot} shows various fluid and field quantities
of our PIC simulation at $t=35$.
They are averaged over $\Delta t=0.25$ to remove noises.
The X-line is located at $(x,z)=(38.1,0.0)$. 
Figure \ref{fig:snapshot}a shows the $x$ component of
the electron bulk velocity $\vec{V}_{e}$.
One can see narrow bi-directional electron jets from the X-line.
The rightward jet ranges $38.1 < x < 48$.
The jet speed is higher than the ion bulk speed $V_{ix}$ and
the inflow Alfv\'{e}n speed $\approx 1.62$ at this time.
This is consistent with previous studies.\citep{prit01a,shay07,kari07}
It has been known that
electrons are unmagnetized in the electron jet region,
while they are magnetized again farther downstream. 
Ref.~\onlinecite{zeni11d} called the jet front boundary ($x\approx 48$) an ``electron shock,'' but
it would be more appropriate to call it the ``remagnetization front.''
In the downstream of the remagnetization front,
unmagnetized ions form a broad current layer.\citep{zeni13,le14}
Properties of this ion current layer was studied in our previous work.\citep{zeni13}
In Figure \ref{fig:snapshot}b,
it is interesting to see
electron divergent flows in the vertical (${\pm}z$) directions
near the remagnetization front.
The maximum jet speed is $|V_{ez}| = 1.67$,
also comparable with the inflow Alfv\'{e}n speed.
These divergent flows generate
the vertical electric currents $J_z$,
which correspond to a step-shaped pattern
in the out-of-plane magnetic field $B_y$
(See Figs.~1c and 3 in Ref.~\onlinecite{zeni13}).
Obviously the super-Alfv\'{e}nic electron jet
is responsible for the divergent flows.
These electron flows further correspond to
narrow electron jets in red ($V_{ex}>0$)
near the separatrices at $x>47$,
as the dashed arrows indicate in Figure \ref{fig:snapshot}a.
These jets penetrate into a broader distribution of
incoming electrons ($V_{ex}<0$).
Hereafter we call these jets the ``field-aligned electron outflows.''
Similar electron jets were reported by recent studies.\citep{zhoum12,zeni13}

Figure \ref{fig:snapshot}c shows the vertical electric fields $E_z$.
They consist of a large-scale X-shaped structure along the separatrices and
a small-scale bipolar structure along the electron jet region.
They are polarization electric fields,
due to a broad ion distribution and a narrow electron distribution. 
Figure \ref{fig:snapshot}d shows
the parallel electric field $E_{\parallel}$.
To better see a weak background structure,
we smooth it with boxcar averaging over ${\sim}0.16$ and then
we adjust the range of the color bar.
Aside from plasma instabilities along the separatrices,
one can recognize a double quadrupole structure.\citep{prit01b,chen09}
The first inner quadrupole is found near the midplane.
It features a strong $E_{\parallel}$ toward the X-line,
a projection of the Hall electric field $E_z$.
This inner quadrupole plays a crucial role for electron dynamics,
as will be shown in Section \ref{sec:noncrossing}.
The second outer quadrupole features
the parallel field $E_{\parallel}$ away from the X-line.
Although they are hard to recognize,
one can see $E_{\parallel}>0$ in the first quadrant ($x \gtrsim 45, z>0$) and
$E_{\parallel}<0$ in the lower half ($x \gtrsim 45, z<0$)
inside the exhaust region.
The other two quadrants are found outside the displayed domain ($x \lesssim 31$).
This quadrupole is a projection of
the reconnection electric field $E_y$ to
the quadrupole Hall magnetic field $B_y$.\citep{sonnerup79}
The incoming electrons near the separatrices,
discussed in the previous paragraph,
are weakly accelerated toward the X-line by this $E_{\parallel}$.

Figure \ref{fig:snapshot}e shows the $x$ component of
the ideal flow vector, $\vec{w} \equiv \vec{E}\times\vec{B}/B^2$.
Although it saturates in the close vicinity of the X-line
($37\lesssim x \lesssim 39, z\approx 0$),
it exhibits a characteristic picture.
It looks bifurcated in the electron jet region:
It is super-Alfv\'{e}nic $w_x = 5$--$6$
on the upper and lower sides of the electron jet.
Strangely,
it is relatively low $w_x = 2$--$3$
at the midplane $z=0$.
It is also enhanced along the separatrices. 
These structures are largely attributed to
the Hall electric field $E_z$ (Fig.~\ref{fig:snapshot}c).
We also note that $w_y$ looks bifurcated
in the electron jet region [not shown].

Figure \ref{fig:snapshot}f shows the electron density $n_e$. 
The ion density looks similar [not shown].
We adjust the color bar $0<n_e<0.3$ for discussion later in this paper, 
while the density reaches $n_e=0.37$ in the downstream side.
At this stage, the reconnection process has flushed out
the Harris current sheet into the outflow region.
The reconnection region is occupied by the inflow plasmas,
whose initial density is $n_{b} = 0.2$.
The typical electron density is $n_e \sim 0.1$--$0.2$ around the center.
One can see two high-density yellow bands around $30<x<45$.
This consists of
a plasma distribution over an ion meandering width ($|z| < 2$--$3$) and
a density cavity near the midplane ($|z| < 1$) in green.
The cavity stretches in $x$ and it covers the electron jet region.

Figure \ref{fig:snapshot}g displays
a nongyrotropy measure $\sqrt{Q}$,
which quantifies the deviation of the electron VDF from gyrotropic one.\citep{swisdak16} 
It is defined in the following way,
\begin{align}
\label{eq:sqrtQ}
\sqrt{Q} \equiv
\Big\{
\frac{P_{e12}^2 + P_{e13}^2 + P_{e23}^2}{P_{e\perp}^2 + 2P_{e\perp}P_{e\parallel}}\Big\}^{1/2},
\end{align}
where
$\overleftrightarrow{P}_{e}$ is the electron pressure tensor.
The subscripts ($\parallel, \perp$) and numeral subscripts indicate
the parallel, perpendicular, and three off-diagonal components of
$\overleftrightarrow{P}_{e}$ in the field-aligned coordinates. 
Equation \eqref{eq:sqrtQ} ranges from $0$ in the fully gyrotropic case
to $1$ in the nongyrotropic limit. 
In Figure \ref{fig:snapshot}g,
the measure highlights the ECL near the midplane.
If one takes a closer look,
two narrow bands are highlighted along the electron jets. 
These are consistent with previous studies.\citep{scudder08,aunai13c,shuster15}
It also marks small-scale regions near the remagnetization front and
the separatrices farther downstream $x > 47$.

Figure \ref{fig:snapshot}h shows
a frame-independent energy dissipation,
$\mathcal{D}_e = \gamma_e [ \vec{J} \cdot (\vec{E}+\vec{V}_e\times\vec{B}) - \rho_c ( \vec{V}_e \cdot \vec{E} ) ],$
where $\gamma_e=[1-(V_e/c)^2]^{-1/2}$ is the Lorentz factor and
$\rho_c$ is the charge density.\citep{zeni11c}
This is equivalent to the nonideal energy conversion
in the nonrelativistic MHD in a neutral plasma.
The measure marks the electron-scale dissipation region around the X-line.
In addition, it also marks the vertical flow region,
as indicated by the circle.

\begin{figure}[tbp]
\centering
\includegraphics[width={0.48\textwidth},clip]{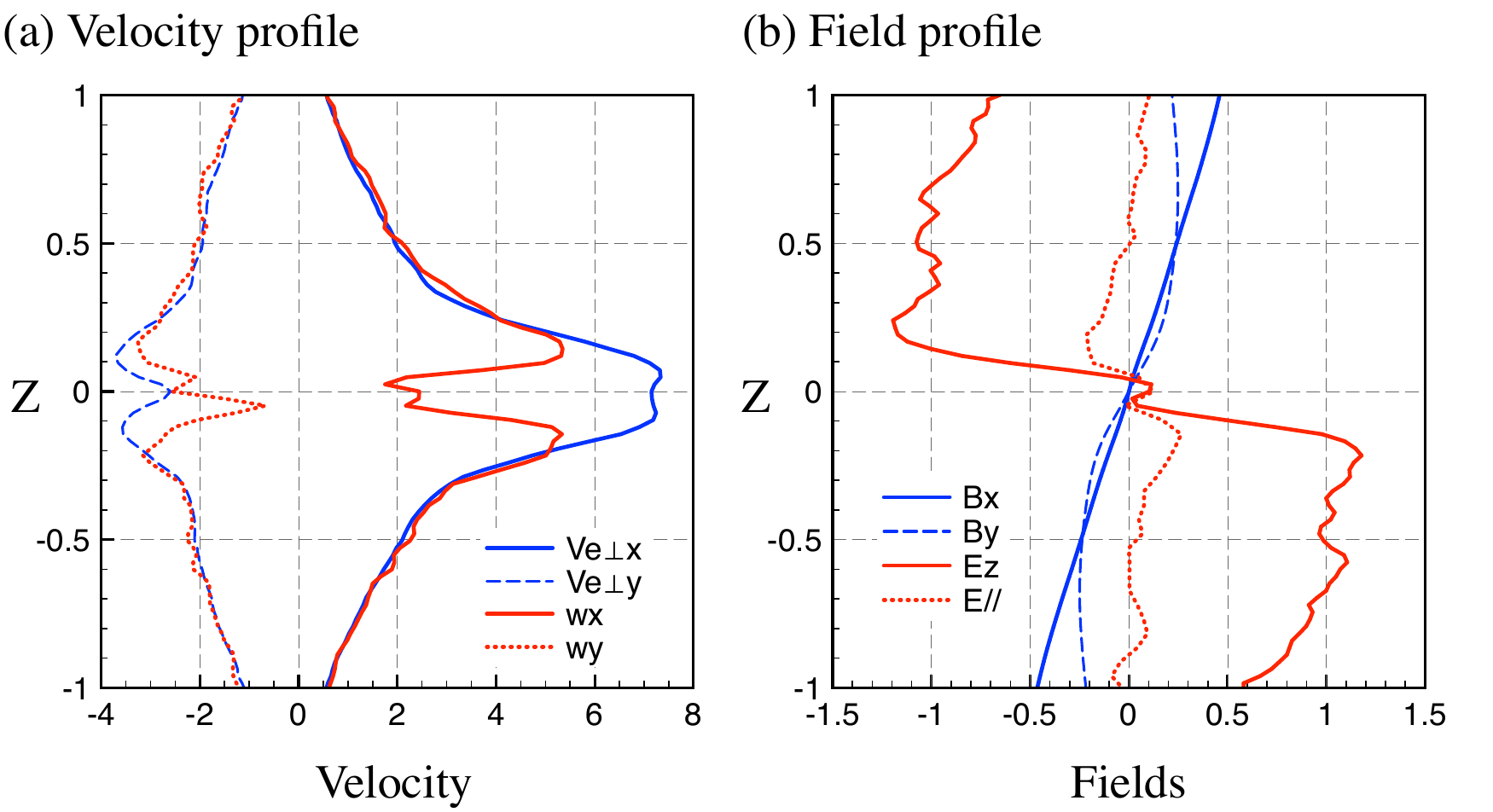}
\caption{(Color online)
1D profiles across the electron jet at $x=43.2$,
averaged over $t=35$--$35.25$.
(a) The electron perpendicular flow $\vec{V}_{e\perp}$ and the ideal flow $\vec{w}$.
(b) The field properties.  The electric field is normalized by $c_{Ai}B_0$.
}
\label{fig:cut432}
\end{figure}

\begin{figure*}[tbp]
\centering
\includegraphics[width={\textwidth},clip]{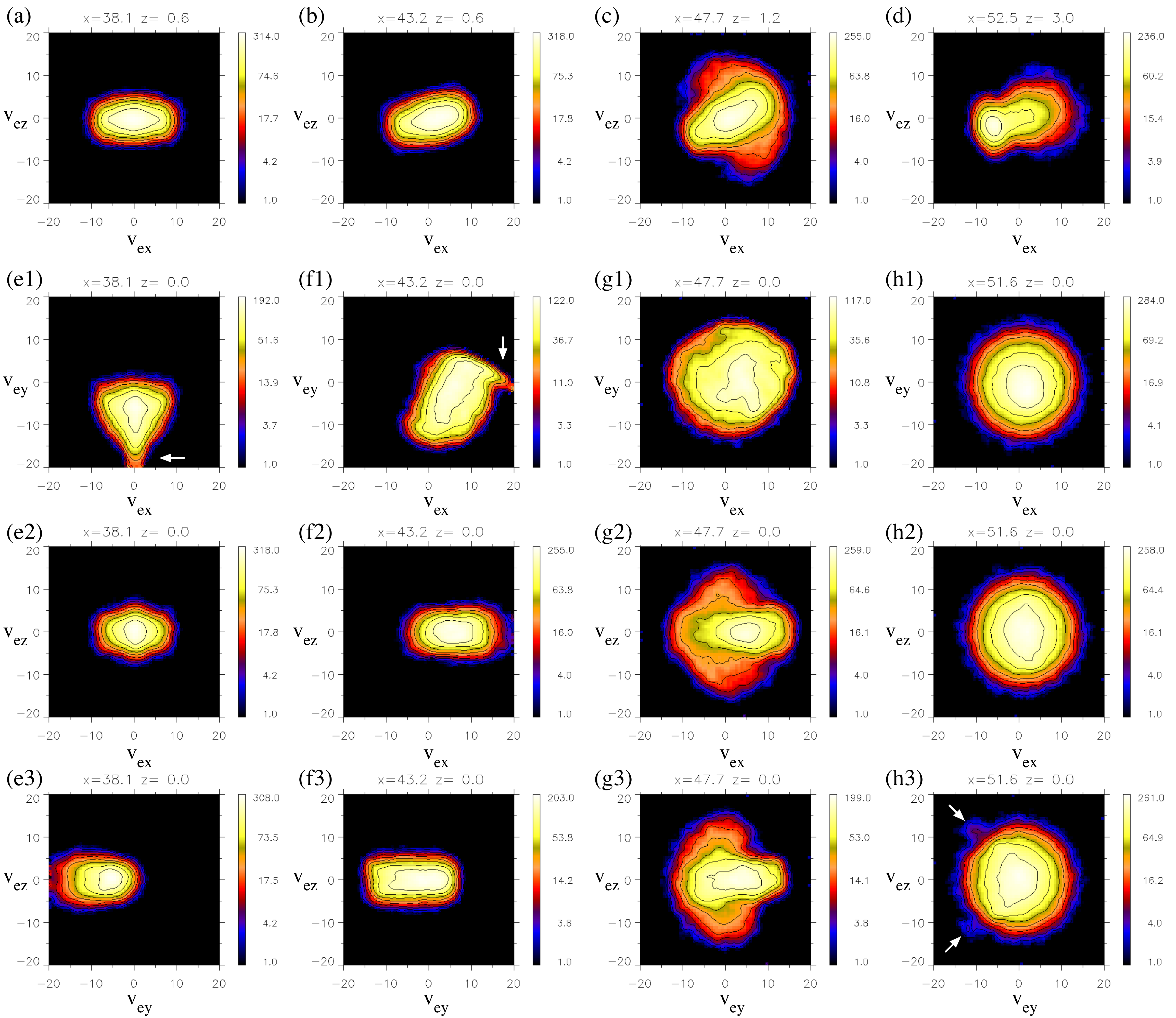}
\caption{(Color online)
Electron velocity distribution functions (VDFs) at $t=35$.
They are computed in a box size of $0.5{\times}0.5$.
The box positions are indicated in Figure \ref{fig:snapshot}a.
For (a-d), one VDF ($v_{x}$--$v_{z}$) is presented.
For (e-h), all three VDFs are presented.
Electron numbers are integrated in the third (out-of-plane) dimension.
}
\label{fig:VDF}
\end{figure*}

Let us take a closer look at the electron jet.
Panels in Figure \ref{fig:cut432} show 1D cuts at $x=43.2$.
The black arrow in Figure \ref{fig:snapshot}a indicates this $x$-position.
The velocity profile (Fig.~\ref{fig:cut432}a) tells us
that the electron perpendicular flow outruns the ideal MHD velocity
($V_{e{\perp}x}\approx V_{ex}>w_x$)
around the ECL ($|z| < 0.2$)
and
that the electron are threaded by the magnetic field
($\vec{V}_{e\perp} \simeq \vec{w}$) outside the ECL. 
The profile of $w_x$ is bifurcated,
as discussed in the previous section. 
The bifurcation of $w_x$ can be seen in Figure 2 in \citet{hesse08} as well. 
The $y$ components ($V_{e{\perp}y}$ and $w_y$) are similarly bifurcated
and $V_{e{\perp}y}$ outruns $w_y$ in the $-y$ direction near the midplane. 
We note that $V_{ex}<V_{e{\perp}x}$ and $V_{ey}<V_{e{\perp}y}$ outside the ECL,
because there is a field-aligned electron outflow
in the $(-x,+y)$ direction toward the X-line.

Figure \ref{fig:cut432}b displays
the variation in the field properties.
Both the reconnecting magnetic field $B_x$ and
the Hall magnetic field $B_y$ change their polarities
across the midplane.
The Hall electric field $E_z$ has a large-scale bipolar structure.
It is negative in $z>0$ and positive in $z<0$.
Its amplitude is eight times stronger than
the reconnection electric field $|E_y|=0.15$.
This also corresponds to
the bipolar ${E}_{\parallel}$ at $|z| \lesssim 0.5$.
Note that parallel electric field remains nonzero in any moving frame,
because $\vec{E}\cdot\vec{B}$ is invariant. 
Outside there,
since ${E}_{\parallel} \simeq 0$ and since $\vec{V}_{e\perp} \simeq \vec{w}$,
the electron ideal condition is recovered,
$\vec{E} + \vec{V}_e \times \vec{B} \simeq 0$.
In a close vicinity of the midplane, $|z| < 0.1$,
one can recognize a reverse bipolar structure in $E_z$ and $E_{\parallel}$.
It is positive in $z>0$ and negative in $z<0$.
This is an electrostatic field
due to the electron meandering motion in $z$.\citep{chen11}


\section{Kinetic Signatures}
\label{sec:kinetic}

\begin{figure}[tbp]
\centering
\includegraphics[width={0.48\textwidth},clip]{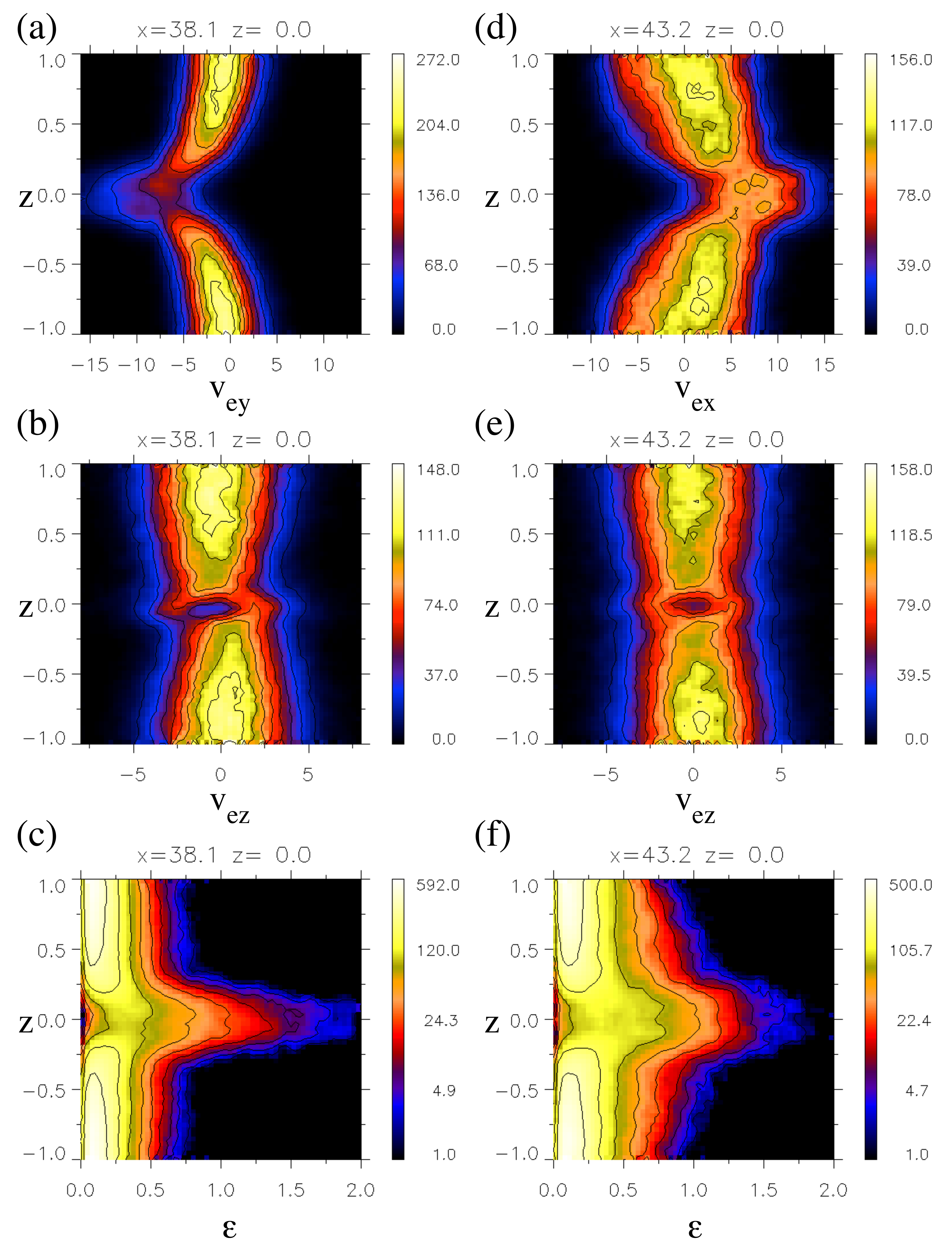}
\caption{(Color online)
Phase-space and energy-space distributions at $t=35$.
The left panels are
(a) the $v_y$--$z$ distribution,
(b) the $v_z$--$z$ distribution, and
(c) the $\mathcal{E}$--$z$ distribution across the X-line ($x=38.1$).
The right panels are
(d) the $v_x$--$z$ distribution,
(e) the $v_z$--$z$ distribution, and
(f) the $\mathcal{E}$--$z$ distribution across the electron jet region ($x=43.2$).
}
\label{fig:phase432}
\end{figure}

Panels in Figure \ref{fig:VDF} show
electron velocity distribution functions (VDFs)
at various locations at $t=35$.
These VDFs are computed in small boxes of $0.5{\times}0.5$.
The boxes are indicated in Figure \ref{fig:snapshot}a and
the figure titles indicate the box center positions.
The VDFs are two-dimensional.
The electron number is integrated in the third direction.

The top four panels (Figs.~\ref{fig:VDF}a--d) display
VDFs ($v_{x}$--$v_{z}$) around the upper boundary regions.
Note that we use lowercase $\vec{v}$ for the particle velocity
in order to distinguish it from the bulk velocity $\vec{V}$.
Figure \ref{fig:VDF}a shows the electron VDF
in $v_{x}$--$v_{z}$ just above the X-line.
It looks highly anisotropic.
Here, the magnetic field is directed in $x$, and so
the electrons are heated in the parallel direction.
\citet{egedal05,egedal08} explained
that the electrons are trapped by the parallel potential
and
that they are fast traveling in the field-aligned directions.
Figure \ref{fig:VDF}b is a VDF at $(x,z)=(43.2,0.6)$.
It looks slightly tilted,
because the magnetic field is directed
in a slightly upward direction.
These two VDFs are gyrotropic. 
The nongyrotropy measure does not mark
these regions (Fig.~\ref{fig:snapshot}g). 

The third panel (Fig.~\ref{fig:VDF}c) shows
a VDF at $(x,z)=(47.7,1.2)$.
In addition to the field-aligned component,
one can see a hot outgoing component in the right half. 
A similar VDF was reported by \citet{chen08}
(\#10 in Fig.~4; \S 2.2.2 in Ref.~\onlinecite{chen08}).
The hot component looks partially gyrotropic;
the electrons are found in the $v_{ey}<0$ half in $v_y$ [not shown]. 
As a result, the entire VDF is weakly nongyrotropic,
because two or more components start to mix with each other here. 
The $\sqrt{Q}$ measure weakly marks
this and nearby regions (Fig.~\ref{fig:snapshot}g). 
Interestingly, this hot component suddenly appears here.
We do not find it in the separatrix regions closer to the X-line.
In Figure \ref{fig:VDF}d,
one can see two field-aligned components,
a cold incoming component and a high-energy outgoing component. 
The cold electrons are weakly accelerated
by $E_{\parallel}$ toward the X-line. 
We will discuss the outgoing electrons later in this paper.
Since both populations are gyrotropic, 
the VDF is anistropic but gyrotropic.

Panels in the bottom three rows show
VDFs at four locations at the midplane. 
The VDFs in $v_{x}$--$v_{y}$ (the second row),
in $v_{x}$--$v_{z}$ (the third), and
in $v_{y}$--$v_{z}$ (the bottom row) are presented.
Figure \ref{fig:VDF}e exhibits typical signatures of a VDF in the DR.
Compared with the inflow region (Fig.~\ref{fig:VDF}a),
the VDF is stretched in $-y$,
due to the $y$ acceleration by the reconnection electric field $E_y$.
The VDF in $v_{x}$--$v_{y}$ looks triangular.\citep{ng11,bessho14,shuster15}
Small $v_{x}$ electrons stay longer in the DR
and therefore
they are more accelerated in $-y$.\citep{prit05}
The structure in $v_z$ is not so clear, 
because our box size in $z$ ($\Delta z = 0.5$) is
larger than the electron meandering width.
The electron VDFs in $|z| < 0.1$ are bifurcated in $v_{z}$ [not shown],
as reported by previous studies.

Figure \ref{fig:VDF}f shows the VDF
at $(x,z)=(43.2,0)$ in the middle of the electron jet.
The overall VDF is shifted in $+v_{x}$
in agreement with the fast bulk flow. 
The bulk velocity is $\vec{V}_{e} \approx (+7,-3,0)$.
The electrons are spread in $v_{x}$--$v_{y}$,
while they are confined in $v_{z}$.
At the midplane $z=0$,
the magnetic field is directed in $z$
and so
the electron perpendicular pressure exceeds the parallel pressure.
This VDF is highly nongyrotropic $\sqrt{Q}\approx 0.4$,
as evident in Figure \ref{fig:snapshot}g.
In $v_{x}$--$v_{y}$ (Fig.~\ref{fig:VDF}f1), 
one can see a narrow ridge in the right,
as indicated by the white arrow.
This is related to $y$-accelerated electrons in the DR,
indicated the white arrow in Figure \ref{fig:VDF}e1.
As we depart from the X-line,
the bottom ridge in Figure \ref{fig:VDF}e1
rotates anti-clockwise and then
evolves into the right ridge in Figure \ref{fig:VDF}f1. 
In $v_{y}$--$v_{z}$ (Fig.~\ref{fig:VDF}f3),
the VDF is weakly bifurcated in $v_z$ for $v_{ey}<0$. 

Figure \ref{fig:VDF}g shows the VDF at $(x,z)=(47.7,0)$ in the jet termination region. 
The VDF looks fairly isotropic in $v_{x}$--$v_{y}$.
In the bottom two panels,
the major outgoing component looks
similar to one in Figure \ref{fig:VDF}f.
In addition, one can recognize
a hot low-density component in the $v_x<0$ half.
These electrons come from the downstream region and then
they start to spread in ${\pm}v_{z}$.
This corresponds to the vertical divergent flows
in Figure \ref{fig:snapshot}b.

Figure \ref{fig:VDF}h shows the VDF at $x=51.6$,
downstream of the remagnetization front.
The electrons are isotropic in all three planes.
One can recognize two small peaks in $v_{y}$--$v_{z}$,
as indicated by the arrows in Panel h3. 
Similar components in VDFs were reported by \citet{shuster14}.
These energetic electrons travel backward from
the downstream magnetic island, but
it is not clear how they are accelerated.
This is one of the earliest signals from the downstream region.
To avoid side-effects from the downstream,
we limit our attention to the upstream side, $x \lesssim 51.6$.

Figure \ref{fig:phase432} provides
additional information to electron kinetic physics.
Figure \ref{fig:phase432}a shows the phase-space diagram in $v_{y}$--$z$
along the inflow line at $x=38.1$.
Here, as the electrons travel in ${\pm}z$
from the inflow regions toward the midplane $z=0$,
they start to drift in $-y$ due to the polarization electric field $E_z$.
Once they enter the ECL,
they are accelerated by the reconnection electric field $E_y$
through Speiser motion.\citep{speiser65}
In the $v_{z}$--$z$ diagram (Fig.~\ref{fig:phase432}b),
a circle around the central hole corresponds to
a bounce motion during the Speiser motion.\citep{hori08,chen11}
Interestingly, the electron density is high outside the ECL, $|z|{\gtrsim}0.5$
(Figs.~\ref{fig:phase432}a and \ref{fig:phase432}b).
Figure \ref{fig:phase432}c is the energy-space diagram in $\mathcal{E}$--$z$,
where $\mathcal{E} = \frac{1}{2}m_ev_e^2$ is
the electron kinetic energy, normalized by $m_i c_{Ai}^2$.
High-energy electrons ($\mathcal{E} > 1.0$) are only found inside the ECL, $|z|<0.25$.

The right Panels show similar diagrams for the electron jet region at $x=43.2$.
Figures \ref{fig:phase432}d and \ref{fig:phase432}e are
the $v_{x}$--$z$ and $v_{z}$--$z$ diagrams. 
The electrons are accelerated in $x$ (Fig.~\ref{fig:phase432}d) and
one can see a phase-space hole in Figure \ref{fig:phase432}e.
There are high-density regions outside the ECL, $|z|{\gtrsim}0.5$
(Figs.~\ref{fig:phase432}d and \ref{fig:phase432}e).
In Figure \ref{fig:phase432}f,
although medium-energy electrons ($\mathcal{E} = 0.5$--$1.0$) are distributed wider in $z$,
high-energy electrons ($\mathcal{E} > 1.0$) are confined around the ECL. 

\section{Particle trajectories}
\label{sec:traj}

\begin{figure*}[tb]
\centering
\includegraphics[width={\textwidth},clip]{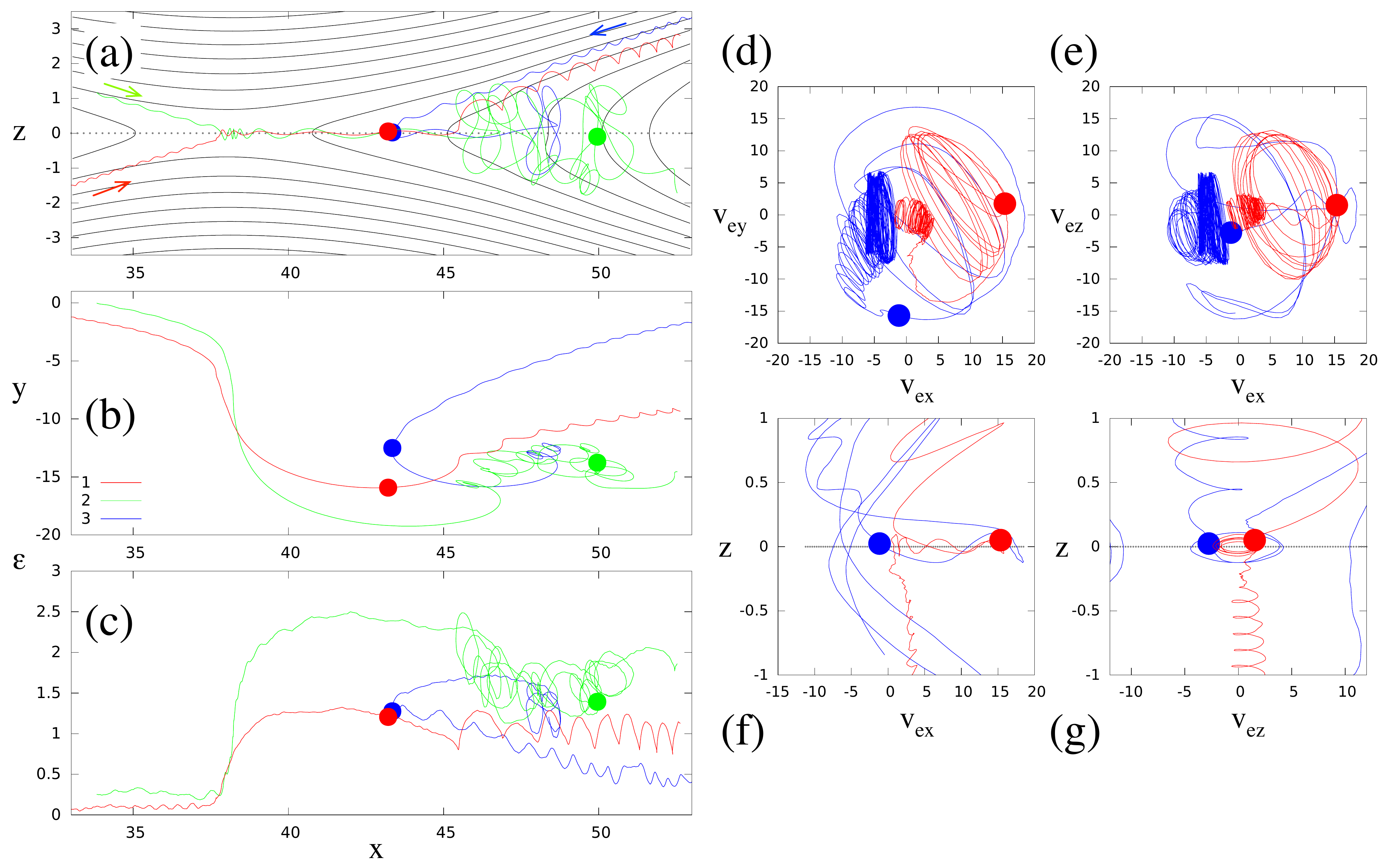}
\caption{(Color online)
Selected electron orbits during $30 < t < 36.5$ in spatial/energy/velocity/phase spaces:
(a) $x$--$z$ plane,
(b) $x$--$y$ plane,
(c) $x$--$\mathcal{E}$ space,
(d) $v_{x}$--$v_{y}$ velocity space,
(e) $v_{x}$--$v_{z}$ velocity space,
(f) $v_{x}$--$z$ phase space, and
(g) $v_{z}$--$z$ phase space.
The right panels focus on the first (red) and the third (blue) electrons.
The circles indicate the particle positions at $t=35$.
The energy $\mathcal{E}$ is normalized by $m_ic_{Ai}^2=m_e(10c_{Ai})^2$ in (c).
}
\label{fig:traj_a}
\end{figure*}

In this work, we manage to record
as many electron trajectories as possible in our PIC simulation. 
Using the particle ID number, we select $3\%$ ($1/32$) of electrons without bias.
Then we output the selected particle data to a hard drive
every two plasma periods $\Delta t = 2\omega_{pe}^{-1}$ during $30<t<36.25$.
The time interval is comparable with one ion gyroperiod, $6.25 \approx 2\pi$.
The time resolution is sufficient to see
electron gyrations, i.e.,
the typical electron gyroperiod is
$2\pi \Omega_{ce}^{-1} \approx 25 \omega_{pe}^{-1} \gg \Delta t$.
It is even sufficient for plasma oscillations around the reconnection site,
i.e., $2\pi\omega_{pe}^{-1}(n_{b}/n_0)^{-1/2} \approx 14\omega_{pe}^{-1} > \Delta t$. 
As a result, we obtain $2.3{\times}10^7$ electron trajectories.
We introduce characteristic trajectories in our dataset
in the following subsections.

\begin{figure*}[t]
\centering
\includegraphics[width={\textwidth},clip]{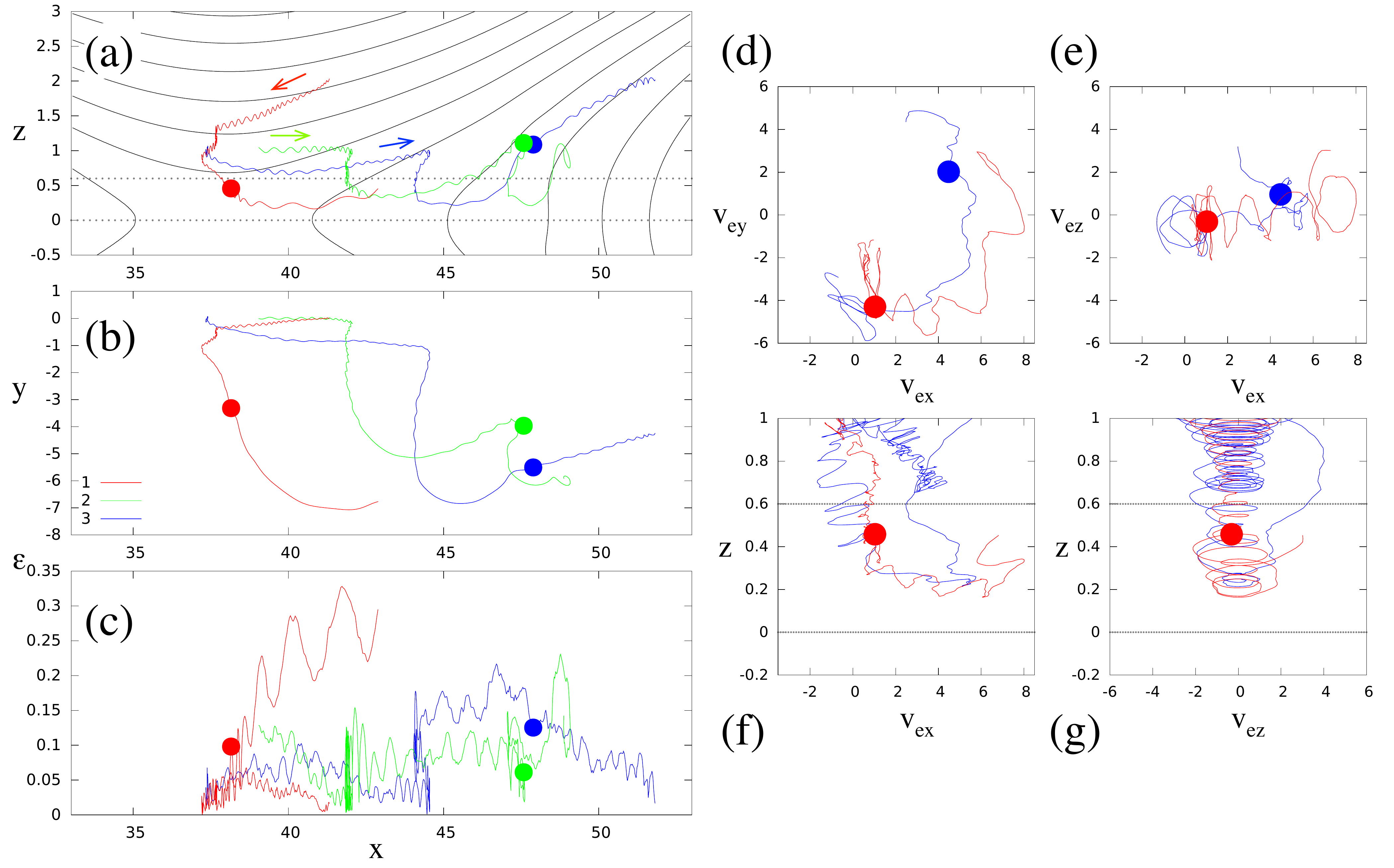}
\caption{(Color online)
Noncrossing Speiser orbits in the same formats as in Figure \ref{fig:traj_a}.
The right panels focus on the first (red) and the third (blue) electrons.
The velocity paths (d,e) are computed in the rage of $0<z<0.6$
between the two dotted lines.
}
\label{fig:traj_b}
\end{figure*}

\subsection{Speiser orbits}
\label{sec:Speiser}

Panels in Figure \ref{fig:traj_a}
show the first set of representative trajectories.
The electron orbits during the interval are presented
in spatial/energy/velocity/phase spaces.
The circles indicate their positions at $t=35$.
The first orbit in red is
a typical Speiser orbit.\citep{speiser65}
In the $x$--$z$ space (Fig.~\ref{fig:traj_a}a),
this electron comes from the bottom left to enter the DR.
Then it is accelerated in the $-y$ direction
by the reconnection electric field $E_y$ (Fig.~\ref{fig:traj_a}b).
As a consequence,
it quickly gains energy near the X-line (Fig.~\ref{fig:traj_a}c).
The electron gradually turns in the $+x$ direction (Fig.~\ref{fig:traj_a}b),
while staying around the midplane ($z \approx 0$; Fig.~\ref{fig:traj_a}a).
Finally it escapes from the midplane to the upper right,
gyrating about the magnetic field line.
In the $v_{x}$--$v_{y}$ space (Fig.~\ref{fig:traj_a}d),
it initially starts from the center,
moves downward in $-v_{y}$ due to the $y$-acceleration near the X-line,
rotates anti-clockwise as it turns in $v_{x}$, and then
exhibits larger gyration after it exits from the midplane.
The $z$-bounce motion around the midplane
is also evident in the central circles
in the $v_{z}$--$z$ space (Fig.~\ref{fig:traj_a}g).

The second orbit in green is
another example of the Speiser orbit.
This one is much more accelerated around the X-line than
the first one (Fig.~\ref{fig:traj_a}c).
After the Speiser rotation,
this electron wanders around the midplane, $|z| \lesssim 1$,
in the downstream region ($x \gtrsim 47$). 
This is interesting, because
we expect that the electron escapes along the field line
like in the first orbit after the Speiser motion.

The third orbit in blue represents
a Speiser motion of different kind.
The electron comes from the upper right
and reaches the midplane at $x\approx 43$ (Fig.~\ref{fig:traj_a}a).
There, it slowly gyrates about $B_z$,
turns its direction from $-y$ to $+x$ while bouncing in $z$.
Instead of passing through the DR,
it is locally reflected to the downstream.
This is a Speiser orbit of local reflection-type. 
Following our previous work on ion orbits,\citep{zeni13}
we call this blue electron orbit a ``local Speiser orbit,'' and
the previous red and green orbits ``global Speiser orbits.'' 
The blue electron gains less energy than
the other electrons through global Speiser orbits.
This is because the local magnetic field $B_z$ is stronger,
and because the electron turns more quickly than in the DR.
During the local reflection phase, 
the velocity vector rotates anti-clockwise
from $-v_{y}$ to $+v_{y}$ (Fig.~\ref{fig:traj_a}d).
This is because
the magnetic field line also turns
from $\pm x$ near the X-line
to $\pm y$ near $x \approx 47$,
where this blue electron escapes from the midplane.
Interestingly, the electron still remains near the midplane,
$|z| \lesssim 1$,
chaotically bouncing in $z$, similar to the second electron in green.

Let us examine the first and third orbits near $x=43.2$ in more detail.
Both are located near the midplane
($|z| \lesssim 0.1$; Figs.~\ref{fig:traj_a}a, \ref{fig:traj_a}f, and \ref{fig:traj_a}g) at $t=35$.
In such close vicinity to the midplane,
both $B_x$ and $B_y$ are approximately linear in $z$ (Fig.~\ref{fig:cut432}b),
while $B_z$ is roughly constant, $B_z \approx 0.06$.
This configuration is similar to the system in Section \ref{sec:theory},
(1) if the system is uniform in $x$ and
(2) if we switch to an appropriately moving frame in which the electric field vanishes.
Even though the two conditions are not exactly met in the PIC simulation,
the theory provides insight into electron motions.
We fit the magnetic field across the midplane at $x=43.2$
to the parabolic model (Eq.~\eqref{eq:B})
to obtain the magnetic curvature radius $R_{\rm c,min}$. 
A similar procedure is presented in Section III B in Ref.~\onlinecite{zeni13}.
At $x=43.2$, the field line is so sharply bent that
the curvature radius is $R_{\rm c,min} = 0.068$.
The electron maximum Larmor radius and
the curvature parameter are as follows,
\begin{align}
\frac{r_{\rm L,max}}{d_i}
&=
\Big( \frac{v'_e}{c_{Ai}} \Big)
\Big( \frac{m_e}{m_i} \Big)
\Big( \frac{B_0}{|B_z|} \Big)
\approx
{1.64}
\Big( \frac{v'_e}{10 c_{Ai}} \Big) \\
\kappa
& \approx
{0.2}
\Big( \frac{v'_e}{10 c_{Ai}} \Big)^{-1/2}
=
{0.16}
\Big( \frac{\mathcal{E}'}{m_i c^2_{Ai}} \Big)^{-1/4}
,
\label{eq:kappa432}
\end{align}
where the prime sign $'$ denotes a physical quantity in a rest frame.
Here we consider an appropriate rest-frame velocity $\vec{U}$,
so that $\vec{v}'_e = \vec{v}_e-\vec{U}$.
In this case,
since the {\bf E}$\times${\bf B} velocity is non-uniform (Fig.~\ref{fig:cut432}a) and
since $E_{\parallel}$ is finite (Fig.~\ref{fig:cut432}b),
it is impossible to find out $\vec{U}$ that transforms away the electric field. 
We approximate $\vec{U}_{43.2}\approx (4,-3,0)$
by referring to the {\bf E}$\times${\bf B} velocity at $z = \pm 0.22$--$0.24$,
where $|E_z|$ hits its maximum (Fig.~\ref{fig:cut432}b).
In this case,
compared with the electron velocities in Figure \ref{fig:traj_a}d,
the frame speed $|\vec{U}_{43.2}|$ is relatively small and so
one can approximate $\mathcal{E}'\approx \mathcal{E}$.
From Equation \eqref{eq:kappa432} and Figure \ref{fig:traj_a}c,
one can see that $\kappa \sim \mathcal{O}(0.1)$ for the two electrons.
This is reasonable, because
the Speiser motion appears in the $\kappa \ll 1$ regime.

Regardless of whether they follow global or local Speiser orbits,
we find that
the electrons undergo either of the following two orbits
after the Speiser phase.
One follows a field-aligned outgoing orbit, like in the first red electron.
This corresponds to the field-aligned electron outflow near the separatrix,
discussed in Section \ref{sec:fluid}. 
The other follows a chaotic bounce motion around the midplane,
as evident in the second green orbit.
It is located at $(x,z)=(50.0,-0.1)$ at $t=35$ and so
we discuss the electron motion near $x \sim 50$. 
This region corresponds the middle of
a broader current layer of unmagnetized ions.\citep{zeni13,le14}
The electric current is weaker, and therefore
the magnetic curvature radius is larger than in the super-Alfv\'{e}nic jet. 
We estimated the curvature radius $R_{\rm c,min}=0.62$ at $x = 50.0$ and
the frame velocity $\vec{U}_{50.0}=(1.37,-0.1,0)$. 
The electric field is excellently transformed away. 
The electron maximum Larmor radius and the curvature parameter are,
\begin{align}
\frac{r_{\rm L,max}}{d_i}
\approx
0.83
\Big( \frac{\mathcal{E}'}{m_i c^2_{Ai}}\Big)^{1/2},
~~
\kappa
\approx
0.86
\Big( \frac{\mathcal{E}'}{m_i c^2_{Ai}}\Big)^{-1/4}
.
\end{align}
One can approximate $\mathcal{E}'\approx\mathcal{E}$,
because the frame speed $|\vec{U}_{50.0}|=1.37$ is negligible.
One can see $\kappa \lesssim 1$ from Figure \ref{fig:traj_a}c.
Figure \ref{fig:traj_a}a tells us that
the typical bounce width $|z|\sim 1$ is comparable with $R_{\rm c,min}=0.62$.
In this regime, the particle motion becomes highly variable. 
Although we do not track the orbit long enough,
the third blue electron exhibits
a similar nongyrotropic motion around $x\gtrsim 48$.

\begin{figure}[btp]
\centering
\includegraphics[width={0.48\textwidth},clip]{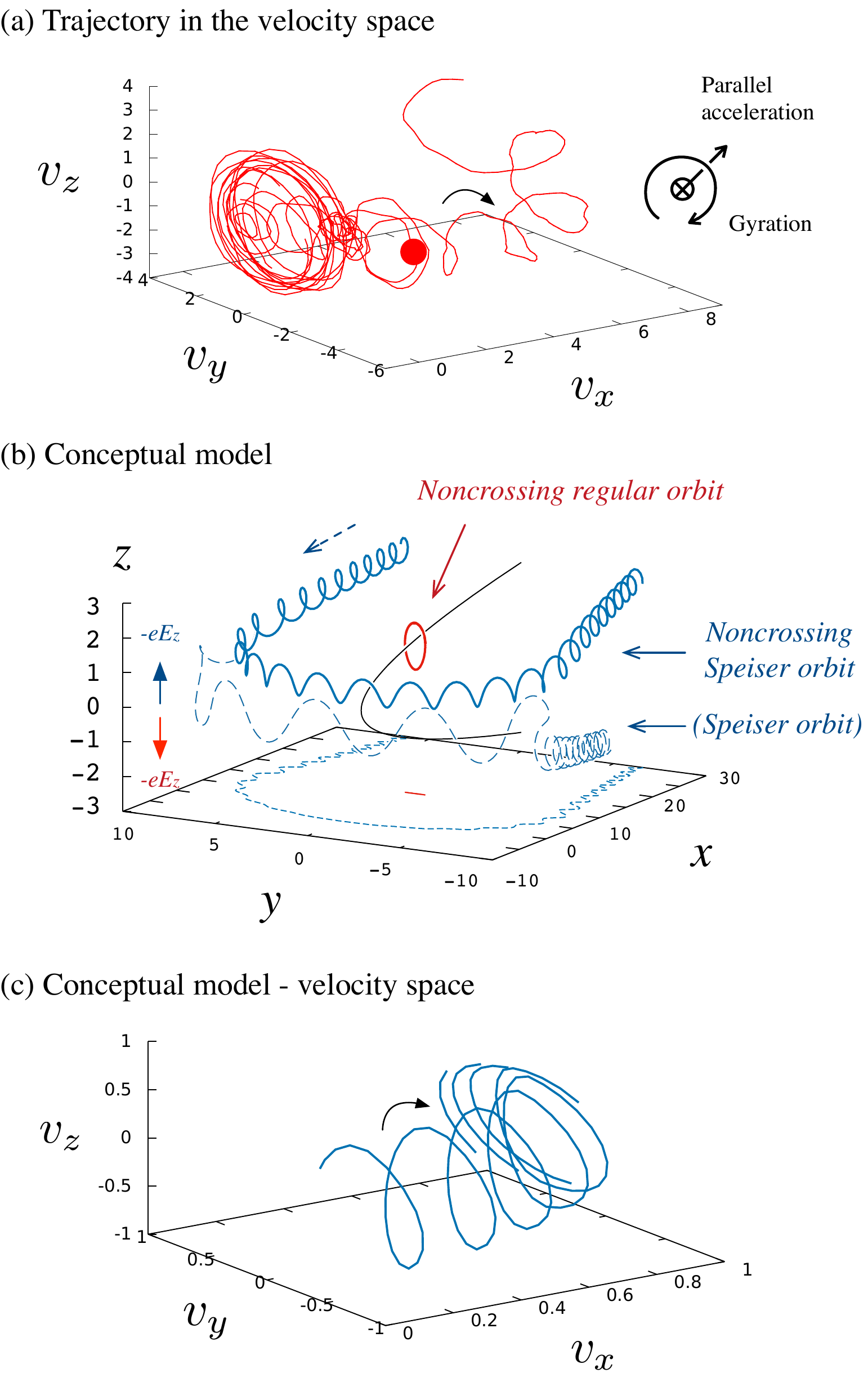}
\caption{(Color online)
(a) Trajectory of the electron \#1 in the 3D velocity space.
(b) A conceptual model of noncrossing electron orbits.
A noncrossing Speiser orbit (solid line) and a noncrossing regular orbit (red).
They are computed in a configuration
similar to the traditional Speiser orbit (dashed line)
in Figure \ref{fig:theory}.
(c) Velocity-space trajectory of the noncrossing Speiser orbit
for $y<0$ and $z<1.5$.
}
\label{fig:orbit247}
\end{figure}

\begin{figure*}[t]
\includegraphics[width={\textwidth},clip]{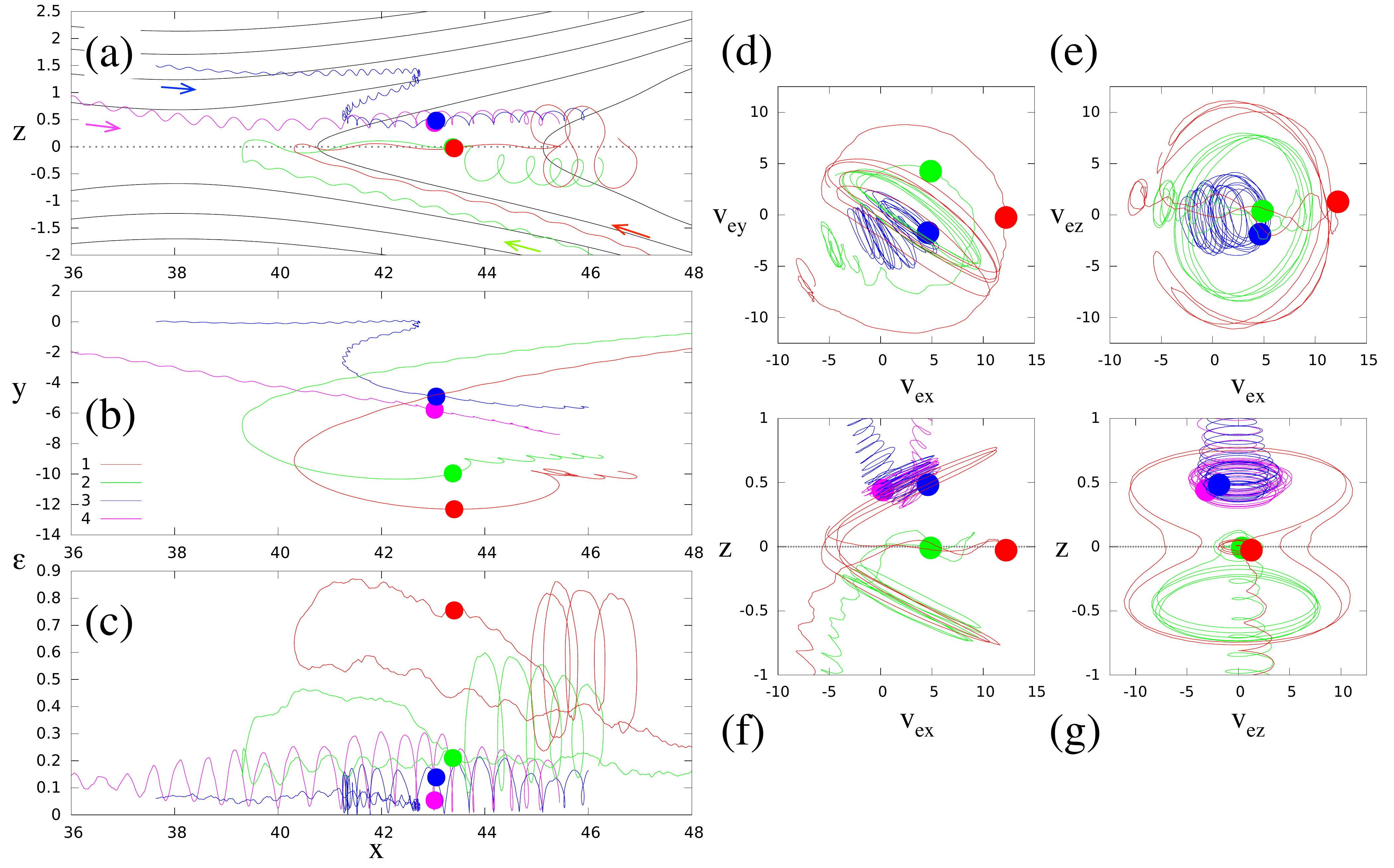}
\caption{(Color online)
Trapped electrons in the same formats as in Figure \ref{fig:traj_a}.
The velocity paths (d,e) are computed in the rage of $-1<z<1$.
The 4th electron in purple is not shown in (d) and (e).
}
\label{fig:traj_c}
\end{figure*}

\subsection{Noncrossing Speiser orbits}
\label{sec:noncrossing}

In Figure \ref{fig:traj_b},
we show a new class of electron orbits in the same format
as Figure \ref{fig:traj_a}.
These electrons approach the midplane $z=0$ and then
all of them exhibit Speiser-like rotations
in the $x$--$y$ plane (Fig.~\ref{fig:traj_b}b).
During the motion, particles bounce in $z$.
However, surprisingly,
they do not cross the midplane (Fig.~\ref{fig:traj_b}a).
To guide our eyes, we indicate $z=0$ and $z=0.6$ by the dotted lines
in Figures \ref{fig:traj_b}a, \ref{fig:traj_b}f, and \ref{fig:traj_b}g.
One can see that all orbits are above the $z=0$ line.
Owing to this, we call them ``noncrossing Speiser orbits.''
Strictly speaking, there is no guarantee that
the reconnecting magnetic field changes its polarity at the midplane, $z=0$.
However, we confirm that the midplane is fairly identical to the field reversal,
because of the perfectly symmetric configuration in $z$ and
a large number of particles per cell in our simulation.
In panels in Figure \ref{fig:snapshot},
the dashed line indicates the field reversal, $B_x=0$.
One can see that it is located at $z \approx 0$ and
that it is sometimes slightly {\em below} the midplane ($z<0$)
near the remagnetization front.
This provides further confidence that
these electrons do not cross the field reversal plane.

First, we examine the motion of electron \#1 in red.
It starts from the upper inflow region (Fig.~\ref{fig:traj_b}a).
Since there is a reconnection electric field $E_y$,
it drifts in $-z$ at the speed of $-E_y/B_x$ while traveling along the field line.
Once it reaches the $0.3 \lesssim z \lesssim 0.8$ region above the DR,
it travels in the $-y$ direction.
Its energy starts to increase (Fig.~\ref{fig:traj_b}c).
The $-y$-motion is attributed to the {\bf E}$\times${\bf B} drift
by the polarization electric field $E_z$.\citep{keizo06,li08}
The gyrocenter velocity approaches $(0,E_z/B_x,-E_y/B_x)$,
which becomes faster in the closer vicinity of the midplane,
because the magnetic field decreases $B_x \rightarrow 0$.
This drift motion in $-y$ is also evident in Figure \ref{fig:phase432}a.
Note that $E_z$ is negative here (Fig.~\ref{fig:snapshot}c).
At $t=35$, the electron is located at $(x,z)=(38.15,0.46)$
with the velocity of $\vec{v}_e = (1.0,-4.3,-0.06)$. 
Below this position, the electron starts to
turn to the outflow direction (Fig.~\ref{fig:traj_b}b)
above the midplane.
Figures \ref{fig:traj_b}d and \ref{fig:traj_b}e present
the velocity-space trajectories within $0<z<0.6$. 
The velocity for the electron \#1 rotates anti-clockwise
in $v_x$--$v_y$ (Fig.~\ref{fig:traj_b}d),
turning to the outflow direction.
During this phase, the electron bounces in $z$
(Figs.~\ref{fig:traj_b}a and \ref{fig:traj_b}e).
These features are similar to those in the Speiser motion.

We further examine the orbit \#1
in the 3D velocity space (Fig.~\ref{fig:orbit247}a). 
One can see that
the velocity vector keeps rotating in the same direction.
This tells us that the electron motion is
a combination of a gyration and a guiding-center motion. 
It is apparently different from
the conventional Speiser motion with a meandering motion,
which exhibits the zigzag pattern in the velocity space (Fig.~\ref{fig:theory}b). 
In Figure~\ref{fig:orbit247}a,
the spiral path further indicates that
the electron is continuously accelerated in the parallel direction. 
We confirm that the electron \#1 is accelerated by a parallel electric field,
by reconstructing the electromagnetic field at the particle position.
Except for minor noises,
the parallel field $E_{\parallel}$ points inward to the X-line,
continuously accelerating the electron away from the X-line in $+x$.
This is consistent with the spatial profile of $E_{\parallel}$ (Fig.~\ref{fig:snapshot}d). 
In the case of the electron \#1,
the parallel acceleration is responsible for
the most of the energy gain, in particular at the later stage. 
On a longer time scale,
the electron velocity slowly rotates anti-clockwise
in $v_x$--$v_y$ (Fig.~\ref{fig:traj_b}d).
One can also interpret that
the electron slowly gyrates about $B_z$,
while the $E_z$ field prevents the particle from crossing beyond a certain distance in $z$.
Note that a field-aligned component $E_{\parallel}$ is
a projection of the polarization electric field $E_z$.
Summarizing these results,
this orbit is similar to but different from the traditional Speiser orbit,
in the sense that
it relies on a combination of the drift motion and the parallel acceleration
instead of the meandering motion. 
Hereafter we call the orbit the ``{\itshape noncrossing} Speiser orbit.''

The second (green) and third (blue) orbits in Figure \ref{fig:traj_b}
are other examples of the noncrossing Speiser orbits.
The electron \#2 exhibits multiple reflections in the $z$ direction.
After entering the central region at $x \approx 42$,
it slowly turns to the $+x$ direction,
travels upward at $x \approx 47$, and then
comes back to the central channel once again.
It travels fast in $x$ and in $-y$ near the midplane
due to the {\bf E}$\times${\bf B} drift by the Hall field $E_z$ (Figs.~\ref{fig:snapshot}c and \ref{fig:snapshot}e),
while it slowly moves in the pedestal region outside the electron jet. 
The electron \#3 in blue travels backward
along the field lines into the central channel at $x=44$,
and then it drifts in the $-y$ direction due to the Hall field $E_z$.
The initial energy of this electron is very low.
It is accelerated to the {\bf E}$\times${\bf B} speed $\approx |E_z/B|$ in this jet flank region. 
Then the electron turns round to $+x$ around $44<x<47$.
One can also see the spiral in the velocity spaces
(Figs.~\ref{fig:traj_b}d and \ref{fig:traj_b}e),
indicating a parallel acceleration by $E_{\parallel}$.
Finally, the electron escapes upward along the field lines. 

We verify the forces acting on
the noncrossing Speiser-orbit electrons
using a conceptual model.
To mimic the Hall field $E_z$,
we impose $\vec{E} = -|v_{e0}|B_0 \sin(\pi z/L) \vec{e}_z$
near the midplane ($|z|<L$) to the parabolic model in Section \ref{sec:theory}.
Here, $|v_{e0}|=1$ is the initial electron velocity outside the Hall-field region ($|z|>L$).
Corresponding electrostatic potential $\int_0^L |E_z| dz = 2/\pi $ is sufficient to
reflect electrons whose normalized energies are $\frac{1}{2}m_e|v_{e0}^2| = 0.5$. 
Figure \ref{fig:orbit247}b displays test-particle orbits in the modified field.
The blue orbit (solid line) employs
the same initial condition as
the Speiser-orbit electron (dashed line)
in Figure \ref{fig:theory}a.
As can be seen,
it excellently reproduces qualitative features for noncrossing Speiser orbits.
The electron remains on the upper half due to the electric field,
turns its direction, and then exits in the $+x$ direction. 
Figure \ref{fig:orbit247}c shows the velocity-space trajectory,
when the electron is in the right half ($y<0$) around the midplane ($z<1.5$).
It exhibits a similar spiral of the {\bf E}$\times${\bf B} drift and
the parallel motion
as in Figure \ref{fig:orbit247}a.

We note that
these noncrossing electrons have lower energy than
the Speiser electrons in Section \ref{sec:Speiser}.
If electrons have enough energies,
they will cross the midplane.
Among the three,
the first one gains the highest energy, 
probably because $B_z$ is weak near the X-line.
Its radius in the $x$--$y$ plane (Fig.~\ref{fig:traj_b}b) is the largest.
The other two are picked up by the outflow exhaust
and then they are locally reflected above the midplane. 
All these features are similar to the Speiser orbits,
even though electrons are always reflected upward
by the Hall field $E_z$ or by the parallel electric field $E_{\parallel}$
above the midplane at $z = 0.2$--$0.4$
(Figs.~\ref{fig:snapshot}c and \ref{fig:cut432}b).
In analogy with the conventional Speiser orbits,
we classify the first one as the noncrossing global Speiser orbit,
and the other two as the noncrossing local Speiser orbits.

\subsection{Regular orbits}
\label{sec:trapped}

Figure \ref{fig:traj_c} presents electron orbits of another kind
in the same format as in Figure \ref{fig:traj_a}.
The first one in red originally comes from the bottom right
and then undergoes the local Speiser motion.
The velocity vector rotates anti-clockwise in $v_{x}$--$v_{y}$ (Fig.~\ref{fig:traj_c}d) and then
the electron eventually turns in the $-x$ direction.
Very interestingly, it starts to bounce in $z$ at $x>45$.
This orbit looks stable.
We argue that this is a regular orbit
in a curved magnetic geometry (Section \ref{sec:theory}). 
The diagonal oscillation in $v_{x}$--$v_{y}$ (Fig.~\ref{fig:traj_c}d),
the inverse C-shaped oscillation in $v_{x}$--$v_{z}$ (Fig.~\ref{fig:traj_c}e), and
the V-shaped path in the phase space (Fig.~\ref{fig:traj_c}f) suggest
a trapped motion in an appropriately moving frame. 
The diagonal oscillation is transverse to
the magnetic fields outside the electron current layer.
The characteristic closed circuit in the $v_z$--$z$ space (Fig.~\ref{fig:traj_c}g)
is consistent with the regular orbit in Figure \ref{fig:theory}d (the red orbit).
At $x=45.6$,
the curvature radius is $R_{\rm c,min}=0.079$ and
the normal magnetic field is $B_z=0.069$.
The corresponding curvature parameter is
\begin{align}
\label{eq:kappa456}
\kappa
& \approx
{0.23}
\Big( \frac{v'_e}{10 c_{Ai}} \Big)^{-1/2}
=
{0.20}
\Big( \frac{\mathcal{E}'}{m_i c^2_{Ai}} \Big)^{-1/4}
.
\end{align}
In this case, it is difficult to estimate the reference-frame velocity $\vec{U}$,
because the ideal velocity $\vec{w}$ has the variation in $z$ (Fig.~\ref{fig:snapshot}e).
We roughly evaluate $v'_e=7.5$--$10$ (Fig.~\ref{fig:traj_c}e) and
$\mathcal{E}' \approx 0.7$ (Fig.~\ref{fig:traj_c}c),
and then obtain $\kappa \approx 0.2$.

In the second case,
the green electron enters the DR and then
undergoes a global Speiser motion.
After leaving the midplane at $x=43.7$,
it starts gyrating in the lower half.
The orbit looks similar to the first regular orbit in red
in the phase spaces (Figs.~\ref{fig:traj_c}f and \ref{fig:traj_c}g),
except that the electron always remains below the midplane.
Figures \ref{fig:snapshot}c--\ref{fig:snapshot}d suggest that
the Hall field $E_z$ keeps the electron away from the midplane. 
Without $E_z$ or $E_{\parallel}$,
an electron usually crosses the midplane in such a field configuration,
because it is reflected by the mirror force toward the midplane.
This electron keeps gyrating around $z \sim -0.5$,
because it is also mediated by the Hall field $E_z$. 
We argue that this is a noncrossing variant of the electron regular orbit.
It is detached from the midplane, due to the Hall field $E_z$.

Both the third electron in blue and the last electron in magenta
travel through similar stable orbits.
The blue one comes from the inflow region and then
enters the stable channel after crossing the separatrix.
The magenta one directly enters the channel,
traveling above the X-line. 
They keep gyrating on the upper flank of the electron jet region
(Figs.~\ref{fig:traj_c}a, \ref{fig:traj_c}f, and \ref{fig:traj_c}g).
They have lower energies than the previous two cases. 
One may interpret these electrons as just drifting.
This is a good point, but we remark that
drift motions have no influence in the parallel motion.
These electrons are trapped in the parallel direction,
balanced by the mirror force toward the midplane
and
the parallel electric force $-eE_{\parallel}$ away from the midplane.
We verify the forces on these orbits
using the test-particle model in Section \ref{sec:noncrossing}.
A stable orbit is shown in red in Figure \ref{fig:orbit247}b.
Therefore, it is appropriate to call the orbits
the noncrossing regular motions, rather than drift motions.

Theoretically, the figure-eight-shaped regular orbits exist
in the field reversal for $\kappa \lesssim 0.53$.\citep{wang94} 
However, we find the figure-eight-shaped regular orbits only for $\kappa \lesssim 0.2$. 
We attribute to this to the Hall electric field $E_z$,
which remains finite in the moving frame.
Keeping electrons away from the midplane,
$E_z$ delays the $z$-bounce motion. 
Recalling that $\kappa$ is the frequency ratio of
the gyration about $B_z$ to the $z$-bounce motion,
the Hall field $E_z$ increases an effective $\kappa$ and
therefore the threshold is reduced to $\kappa \approx 0.2$.
One can also interpret in the following way:
While lower-$\kappa$ (higher-energy) electrons are insensitive to $E_z$,
higher-$\kappa$ (lower-energy) electrons are sensitive.
The Hall field $E_z$ transforms
the high-$\kappa$ figure-eight-shaped orbits ($0.2 \lesssim \kappa \lesssim 0.5$)
to the noncrossing regular orbits.
In Figures \ref{fig:traj_c}c--\ref{fig:traj_c}e,
one can see that
all the noncrossing regular electrons
have lower energy than
the crossing electron in red
in an appropriate frame. 
Although $\kappa$ is not so meaningful for the noncrossing electrons, 
we plug in their typical velocities to Eq.~\ref{eq:kappa456}
to obtain $0.2 \lesssim \kappa \lesssim 0.4$. 
This further suggests that
they are detached variants of the regular orbits.

\begin{figure*}[tbp]
\centering
\includegraphics[width={\textwidth},clip]{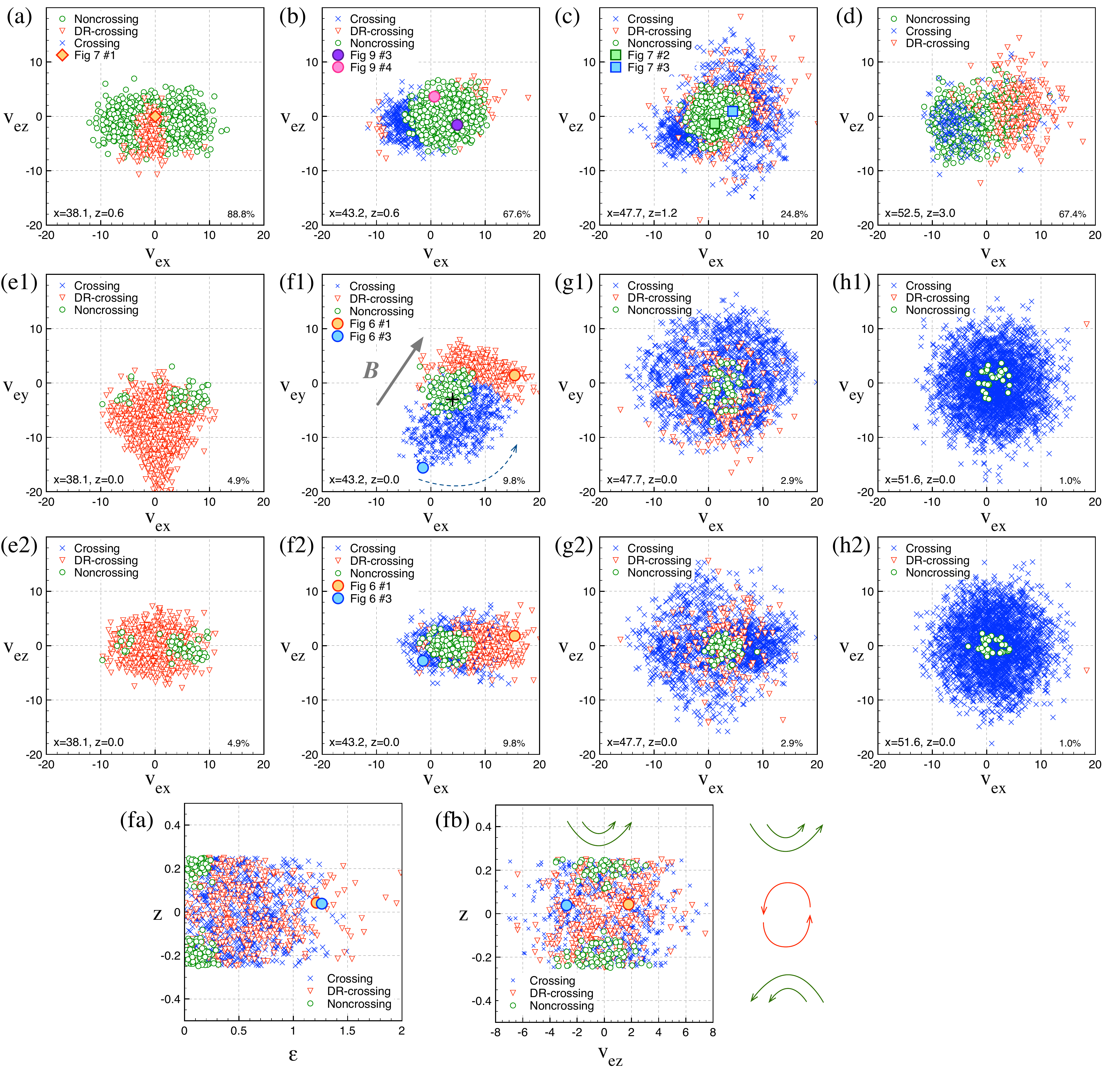}
\caption{(Color online)
Composition of electron velocity distribution functions (VDFs) at $t=35$.
They are computed in a box size of $0.5{\times}0.5$.
The box positions are indicated in Figure \ref{fig:snapshot}a.
(a-h) Electron velocity distribution functions (VDFs)
in the same format as in Figure \ref{fig:VDF}.
(fa) Energy-space distribution in $\mathcal{E}$--$z$ (Fig.~\ref{fig:phase432}f),
and
(fb) phase-space distribution in $v_{z}$--$z$ (Fig.~\ref{fig:phase432}e),
computed in the (f) region ($x,z \in [42.95, 43.45] \times [-0.25, 0.25]$).
The electrons are classified as
noncrossing electrons (green circles),
DR-crossing electrons (red triangles),
and
other noncrossing electrons (blue crosses).
See the text for further detail.
}
\label{fig:VDF2}
\end{figure*}

\section{Composition Analysis}
\label{sec:comp}

The trajectory dataset allows us to
explore kinetic signatures in further detail.
Figure \ref{fig:VDF2} shows the velocity,
energy, and phase-space distribution of electrons at $t=35$.
Each symbol stands for the electrons in the dataset.
Based on their trajectories,
we classify the electrons into the following three classes.
The green circles indicate
electrons that never cross the midplane $z=0$
during the interval ($30<t<36.25$).
We call them ``noncrossing candidates.''
They may cross the midplane
before $t=30$ or after $t=36.25$. 
The red triangles indicate electrons that
spent some time in the square region of $x,z \in [35.5,40.5] \times [-0.2, 0.2]$
during the interval.
This region approximates the DR,
which is indicated in Figure \ref{fig:snapshot}h.
We call them ``DR-crossing electrons.'' 
The blue crosses indicate the other electrons.
They have crossed the midplane at least once,
but they spent no time in the DR  during the interval.
We call them ``crossing electrons.''
Please note that
the frontmost symbols sometimes overwrite the background ones
in panels in Figure \ref{fig:VDF2}.
The order is carefully selected to
emphasize interesting features.

Panels in the first three rows show
the electron VDFs at $t=35$.
They are equivalent to those in Figure \ref{fig:VDF}. 
Typical features of the separatrix VDFs
are found in Figures \ref{fig:VDF2}b--d.
In general, one can recognize
outgoing red particles from the DR and
incoming blue electrons toward the midplane.
In Figure \ref{fig:VDF2}b,
the leftmost electrons turn red,
because they are going to enter the DR.
Figure \ref{fig:VDF2}c contains the outgoing blue population,
which crossed the midplane outside the DR.
In Figure \ref{fig:VDF2}d,
the outgoing red population is more prominent in the $v_{ex}>0$ half,
even though the outgoing blue population is also hidden behind them.
This is reasonable because
the electrons from the DR (global Speiser electrons) are
more energetic than the local Speiser electrons, and therefore
the DR-crossing electrons travel deeper into the exhaust region
beyond the remagnetization front.

Surprisingly, we find a substantial amount of noncrossing candidates
in the first four VDFs (Figs.~\ref{fig:VDF2}a--d). 
In these panels, a number on the bottom-right corner indicates
the ratio of the number of noncrossing candidates to the total number.
In Figure \ref{fig:VDF2}a, right above the DR,
89\% of the electrons are noncrossing candidates.
The noncrossing candidates are also hidden
behind the central red population.
For example, the diamond symbol indicates
the electron \#1 in Figure \ref{fig:traj_b}.
As discussed, it travels through the global noncrossing Speiser orbit. 
Here this electron is classified as noncrossing electrons in green. 
Interestingly, this electron also hits
the DR of $x,z \in [35.5,40.5] \times [-0.2, 0.2]$,
even though it does not cross the midplane. 
In contrast, only a limited number of electrons
are entering the DR and crossing the midplane. 
The red population is found only around the center $|v_{ex}| \sim 0$,
while left-going and right-going populations are noncrossing.
In the next domain (Fig.~\ref{fig:VDF2}b),
although some blue crosses are hidden behind the green circles,
the noncrossing candidates are majority,
accounting for 68\% of the total electron number.
The purple and magenta circles indicate
the electrons \#3 and \#4 in Figure \ref{fig:traj_c}.
As discussed in Section \ref{sec:trapped},
they are trapped on the upper flank of the electron jet,
traveling through noncrossing regular orbits.
The relevant blue orbit in $v_x$--$v_z$ (Fig.~\ref{fig:traj_c}e) is
in excellent agreement with the green region in Figure \ref{fig:VDF2}b.
These results suggest that
the green noncrossing population in Figure \ref{fig:VDF2}b
are likely to travel through the noncrossing regular orbits. 

Farther away from the DR,
in Figure \ref{fig:VDF2}c,
25\% of electrons are noncrossing candidates. 
The two squares indicate
the noncrossing Speiser electrons \#2 and \#3
in Figure \ref{fig:traj_b}.
They are either reflecting back to
the midplane (\#2) or escaping outward (\#3).
In this VDF, the noncrossing candidates are found around the center.
Their velocity distribution is fairly unchanged
from the previous case (Fig.~\ref{fig:VDF2}b). 
The hot outgoing population
consists of the crossing electrons in either red or blue.
This suggests that they are Speiser-accelerated electrons from the midplane.
One can see in Figure \ref{fig:traj_a} that
the Speiser-accelerated electron in red wraps around the magnetic field line  along the separatrix (Fig.~\ref{fig:traj_a}a).
Its velocity (Fig.~\ref{fig:traj_a}e) explains
the hot population in Figure \ref{fig:VDF2}c very well.
The hot population is evident
in the (c) region and in further downstream along the separatrix,
because Speiser electrons are ejected from the midplane at the end of the ECL,
as can be seen in the orbit \#3 (blue) in Figure \ref{fig:traj_a}a
(See also Fig.~4(a) in Ref.~\onlinecite{chen11}).
This agrees with
the divergent flows in ${\pm}z$ (Fig.~\ref{fig:snapshot}b; Sec.~\ref{sec:fluid})
and
the vertically spread VDF (Fig.~\ref{fig:VDF}g; Sec.~\ref{sec:kinetic}).

In Figure \ref{fig:VDF2}d,
we recognize green noncrossing candidates in the incoming direction ($v_x<0$).
Some are hidden behind the blue crossing electrons.
These green candidates may be overemphasized,
because the (d) region is far from the DR and the midplane.
The field-line through $(x,z)=(52.5,3.0)$ at $t=35$ is convected to
$(x,z)=(48.4,0.0)$ at $t=36.25$,
and
the field-aligned distance to the midplane is $\approx 5 [d_i]$.
The electrons at a velocity $|v_e|=4$
will travel $5 [d_i]$ from $t=35$ to $t=36.25$.
Therefore, some green electrons could be crossing electrons.
On the other hand,
our classification will be valid in the outgoing part ($v_x>0$).
One can see energetic electrons from the DR in red.
They are more pronounced than blue crossing populations.
Importantly,
the green population retains signatures similar to
those in the previous cases.
From these four panels,
we find a non-negligible amount of noncrossing electrons.
This will be further analyzed in this section.

Panels in the second and third rows show the VDFs at the midplane,
similar to those in Figure \ref{fig:VDF}.
The left panels (Fig.~\ref{fig:VDF2}e) indicate
the VDFs around the X-line.
We recognize some amount of green noncrossing electrons here,
because the VDF is calculated in a thicker box in $z$ than
the square region to classify the red population,
and because some noncrossing electrons come close to the midplane
(e.g., the orbit \#1 in Fig.~\ref{fig:traj_b}a).
Aside from them,
the (e) region is filled with the DR-crossing electrons in red.
As we depart from the X-line in the $+x$ direction,
the blue population gradually replaces the red population in the VDFs.
In the $v_x$--$v_y$ space (Fig.~\ref{fig:VDF}f1),
the blue population appears in the bottom ($v_{ey}<0$).
Then they evolve anti-clockwise,
as the dashed arrow indicates.
The red population rotates anti-clockwise accordingly. 
Finally, all these electrons are mixed with each other
around the remagnetization front (Fig.~\ref{fig:VDF}g).
We see no remarkable separation in color farther downstream.

In Figure \ref{fig:VDF2}f1,
we argue that
the global Speiser motion accounts for the DR-crossing electrons in red
and
that the local Speiser motion accounts for the other crossing electrons in blue.
The two circles in Figure \ref{fig:VDF2}f indicate
the representative electrons
for the global Speiser motion
(the orbit \#1 in Fig.~\ref{fig:traj_a}) and
for the local Speiser motion (\#3),
discussed in Section \ref{sec:Speiser}.
For the local-type Speiser motion,
we expect a half-ring distribution function in $v_x$--$v_y$,
corresponding to the slow half-gyration about $B_z$.
In Figure \ref{fig:VDF2}f1,
the gray arrow indicates
the orientation of the magnetic field at $z = 0.22$--$0.24$. 
It is tilted by 56 degrees due to the Hall effect. 
In Section \ref{sec:Speiser},
we estimated the frame velocity $\vec{U}_{43.2}=(4,-3,0)$.
This is indicated by the black cross in Figure \ref{fig:VDF2}f1.
Keeping these in mind,
one can see that the blue electrons are distributed in
a semicircle or a half ring surrounding $\vec{U}_{43.2}$
in this velocity space.
The semicircle is tilted,
similar to the magnetic field outside the ECL (the gray arrow). 
From Figures \ref{fig:VDF2}f1 and \ref{fig:VDF2}f2,
one can see the typical velocity for the blue electrons
${v}'_e=|\vec{v}_e-\vec{U}_{43.2}| \approx 5$--$15~c_{Ai}$,
which corresponds to the Speiser regime of $\kappa < 1$ (Eq.~\ref{eq:kappa432}). 
All these features are consistent with
the Speiser motion of local-reflection type. 

In the (f) region,
the green noncrossing candidates are found
near $\vec{U}_{43.2}$ in the velocity spaces.
Their thermal velocity is smaller than in the upper (b) region,
probably because they lose their energy due to the Hall field $E_z$.
Some more signatures of the green noncrossing electrons are evident
in the energy-space and phase-space diagrams
for the electron distribution (Figs.~\ref{fig:VDF2}fa and \ref{fig:VDF2}fb),
which correspond to Figs.~\ref{fig:phase432}f and \ref{fig:phase432}e.
In contrast to the two crossing populations (blue and red),
the green noncrossing candidates are found
only outside the midplane $|z| \gtrsim 0.1$.
Their energy is low, $\mathcal{E} \lesssim 0.3$,
in agreement with small thermal velocity in the VDFs.
By definition, particles move downward (upward) in the $v_{ez}<0$ ($v_{ez}>0$) region
in the $v_z$--$z$ space (Fig.~\ref{fig:VDF2}fb).
With this in mind, we see that
the green electrons are reflected away from the midplane.
They rotates anti-clockwise near the $v_{z}=0$ axis,
as indicated by the green arrows.
These features are consistent with
the noncrossing orbits in Section \ref{sec:traj}
(see Figs.~\ref{fig:traj_b}g and \ref{fig:traj_c}g).

\begin{figure}[bhtp]
\centering
\includegraphics[width={0.48\textwidth},clip]{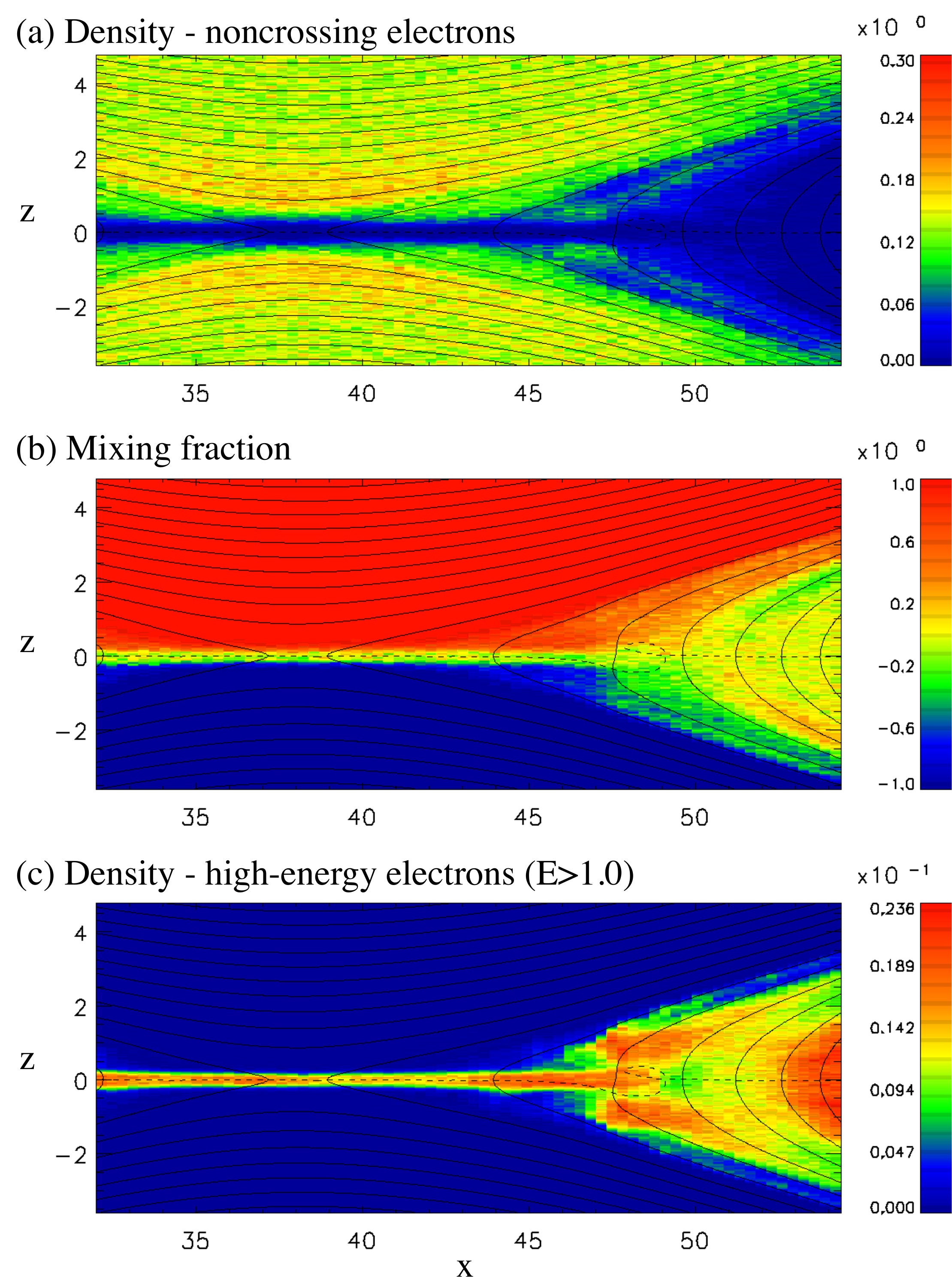}
\caption{(Color online)
(a) Reconstructed density of the noncrossing electrons at $t=35$.
The color range is set similar to Fig.~\ref{fig:snapshot}f.
(b) Mixing fraction $\mathcal{M}$, defined by Eq.~\eqref{eq:mixing}.
(c) Density of the high-energy electrons for $\mathcal{E}>1.0$.
}
\label{fig:dist}
\end{figure}

Next, we investigate
spatial distribution of the noncrossing candidates.
Figure \ref{fig:dist}a shows
the density of the noncrossing candidates,
reconstructed from our 3\% dataset,
in the same color range as in Figure \ref{fig:snapshot}f.
Around the ECL, there are three layers of
(1) the high-density yellow layers in the inflow region,
(2) the medium-density green layers near the separatrices, and
(3) the blue cavity along the midplane.
Comparison between Figures \ref{fig:snapshot}f and \ref{fig:dist}a
tells us that
the noncrossing electrons are the majority in the inflow region,
in particular in the high-density yellow layer.
The noncrossing electron density decreases in the medium-density green layers,
but it is non-negligible $\sim 0.1$.
The noncrossing electrons disappear
in the downstream of the remagnetization front, $x \gtrsim 48$.
We note that
the noncrossing electron density could be underestimated
in the flanks of the ECL at $45 \lesssim x \lesssim 47$.
Even if some electrons do not cross the midplane at $t<35$,
once they cross the midplane somewhere in the downstream ($x \gtrsim 48$) during $35<t<36.25$,
we count them as crossing electrons in our analysis.
Despite these concerns, we recognize many noncrossing candidates.

The presence of noncrossing electrons implies that
upper-origin and lower-origin electrons may not
mix with each other across the ECL.
Figure \ref{fig:dist}b shows
the electron mixing fraction at $t=35$,
computed from the full PIC datasets.
The fraction is defined in the following way
\begin{align}
\label{eq:mixing}
\mathcal{M} \equiv \frac{N_{\rm up}-N_{\rm low}}{N_{\rm up}+N_{\rm low}}
\end{align}
where
$N_{\rm up}$ is the number density of electrons
that were in the upper half ($z>0$) at $t=30$ and
$N_{\rm low}$ in the lower half ($z<0$).
It ranges from
$\mathcal{M} \rightarrow +1$ in the upper inflow region
to
$\mathcal{M} \rightarrow -1$ in the lower region.
During $30<t<35$, the magnetic flux across the X-line is transported
by 1.7 in ${\pm}z$ and by 11.4 in ${\pm}x$. 
The latter is comparable with
the distance between the X-line and the remagnetization front.
Thus, we expect that electrons are fully mixed $\mathcal{M} \approx 0$
in the exhaust region between the separatrices.
However, surprisingly,
the electrons are mixed only inside the ECL
in the upstream of the remagnetization front ($x{\lesssim}48$).
They are largely unmixed outside the ECL. 
Weakly mixed regions around $45 < x <48$
between the ECL and separatrices
do not change this picture.
The electrons are quickly mixed in the downstream, $x{\gtrsim}48$. 
Based on these results,
we conclude that electron mixing is inefficient
in the upstream side of the remagnetization front ($x{\lesssim}48$)
and that
the electron mixing occurs
mainly in the downstream of the remagnetization front.

Crossing electrons are distributed
in the outflow region between the separatrices.
Many of them follow the Speiser orbits.
Through Speiser-type orbits,
electrons can be accelerated to higher energies than the noncrossing electrons.
Motivated by this, we examine
the spatial distribution of energetic electrons.
Figure \ref{fig:dist}c shows a number density of electrons
whose energy exceeds a threshold, $\mathcal{E} > 1.0$.
They are localized around the ECL ($x<48$).
The localization of the high-energy electrons is also
evident in Figures \ref{fig:phase432}c and \ref{fig:phase432}f.
They are crossing populations,
as confirmed in Figure \ref{fig:VDF2}fa.
These electrons follow
either the global Speiser orbits from the DR or
the local Speiser orbits that turn around inside the ECL. 
After the Speiser phase,
these electrons escape along the separatrices or
they chaotically remain around the midplane,
as discussed in Section \ref{sec:Speiser}.
In Figure \ref{fig:dist}c,
the energetic electrons are located on separatrices
in the downstream of the remagnetization front ($47<x<53$).
This supports the former (the orbit \#1 in Fig.~\ref{fig:traj_a}),
while we do not see significant energization near the midplane.
We find that
some nongyrotropic electrons lose their energy
as shown by the orbit \#2 in Figure \ref{fig:traj_a}c. 
In these regions,
\citet{hoshino01a} proposed a two-step mechanism of
the Speiser acceleration and the $\nabla{B}$ acceleration.
Our results are not favorable to the \citet{hoshino01a}'s proposal,
probably because combinations of acceleration mechanisms vary from case to case. 
Farther downstream ($x>52$),
the energetic electrons are again found near the midplane.
The region is equivalent to
an outer edge of a long magnetic island,
inside which these electrons are confined.
We confirm that the energetic electrons are
repeatedly accelerated inside the island
across the periodic boundary in $x$.\citep{drake06}
Since we focus on electrons from the ECL side,
these energetic electrons are out of the scope of this study.

In summary,
a substantial amount of noncrossing electrons are found outside the ECL.
They have less energy than
the high-energy population inside the ECL,
because the crossing electrons are accelerated via the Speiser process,
once they enter the ECL.
The energetic electrons exit from the ECL toward the separatrices. 

\section{Observational signatures}
\label{sec:ET}

\begin{figure}[htbp]
\centering
\includegraphics[width={0.48\textwidth},clip]{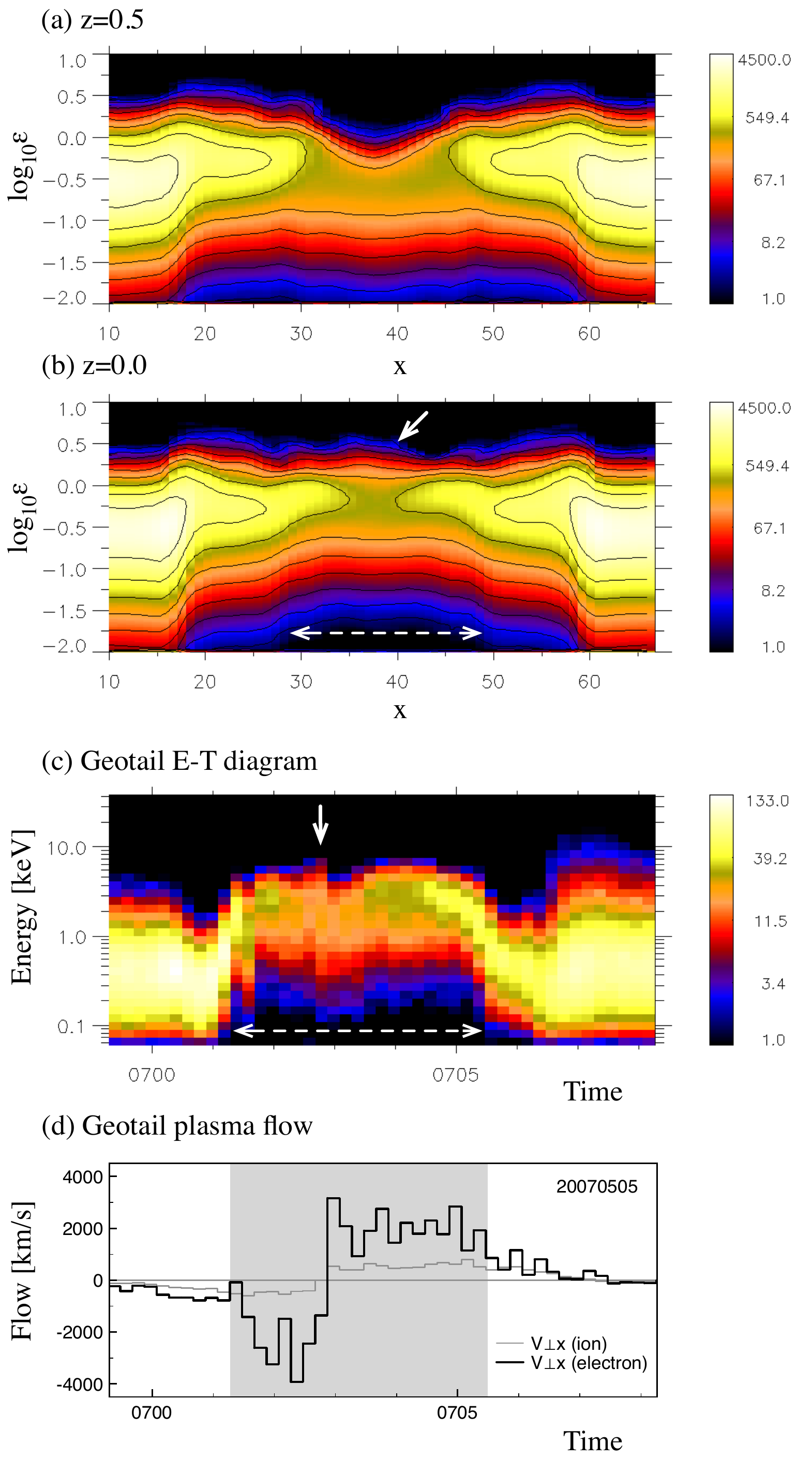}
\caption{(Color online)
Electron energy-space spectrograms at $t=35$ at two slices
(a) $z=0.5$ and (b) $z=0.0$
in the PIC simulation.
(c) Electron energy-time spectrograms and
(d) plasma perpendicular flows,
for the period from 0659:18 UT to 0708:16 UT on 5 May 2007,
observed by the Geotail satellite.\citep{nagai13b}
}
\label{fig:ET}
\end{figure}

We discuss potential observational signatures in this section.
Figures \ref{fig:ET}a and \ref{fig:ET}b show
electron energy-space spectrograms (E-S diagrams),
computed from the PIC simulation at $t=35$. 
We count the electron particle {\em flux}
as a function of the logarithmic energy
above the ECL at $z=0.5$ (Fig.~\ref{fig:ET}a)
and
along the ECL at $z=0.0$ (Fig.~\ref{fig:ET}b).
The spatial resolution is $\Delta x = 0.89, \Delta z =0.5$.
The vertical axis is equivalent to
the energy spectrum of
$\mathcal{E}^{1.5}f(\mathcal{E})d\mathcal{E}$.
In Figure \ref{fig:ET}b,
one can see a two-step profile of the electron count rates.
The reconnection site ($18<x<59$) is filled with
tenuous plasmas from the inflow region. 
The ECL ($28<x<48$) is embedded inside the region,
as indicated by the dashed arrows.
Around the X-line,
we recognize many energetic electrons of $\mathcal{E}>1$ (the solid arrow). 
Since they are absent above the ECL (Fig.~\ref{fig:ET}a),
they are quite probably accelerated near the X-line via the Speiser process.
\citep{speiser65,zeni01,prit05}
The energy spectrum has a power-law tail
$f(\mathcal{E})d\mathcal{E}\propto\mathcal{E}^{-5.8}$
around the X-line.
As we depart from the X-line in the ECL,
the spectral index gradually decreases. 
The electron fluxes slightly shift to the higher energies
and then shift to the lower energies. 
Another remarkable signature is
the absence of the low-energy electron flux in the ECL.
One can see that
the electron flux of $10^{-1.5}$--$10^{-1}$ suddenly disappears
around the ECL in Figure \ref{fig:ET}b.
In contrast, the low-energy electron flux remains fairly unchanged
above the ECL (Fig.~\ref{fig:ET}a).

The bottom two panels in Figure \ref{fig:ET}
show Geotail observation of magnetic reconnection
from 0659:18 UT to 0708:16 UT on 5 May 2007.
The satellite was located in the magnetotail at (-21.3, 6.9, 1.3 $R_E$)
in the geocentric solar magnetospheric (GSM) coordinates
at 0700 UT. 
This event was studied by \citet{nagai13b} in detail.
Here we briefly introduce key signatures.
Figure \ref{fig:ET}d presents the ion and electron bulk velocities,
obtained from plasma moments.
The $x$ components of plasma perpendicular velocities are presented.
Both the ion velocity ($V_{{\perp}x}$ in gray color) and
the electron velocity ($V_{{\perp}x}$ in black) are initially negative,
reverse their signs at 0702:41 UT, and then
remain positive until $\sim$0707 UT.
This and other signatures\citep{nagai13b} suggest that
the Geotail encountered bidirectional reconnection outflows
in the close vicinity of the X-line. 
In the period 0701:17--0705:29 UT, 
the electron flow is decoupled from the ion flow.
The shadow in Figure \ref{fig:ET}d indicates
this ion-electron decoupling interval. 
Figure \ref{fig:ET}c shows electron counts per sample time
in the energy-time (E-T) diagram.
If the structure of the reconnection region is stationary,
the E-S diagrams are equivalent to the E-T diagram. 
One can see that
the electron fluxes shift to higher energies
during the reconnection event from $\sim$0700 UT to $\sim$0707 UT.
The dashed arrows indicated
the ion-electron decoupling interval (0701:17 UT to 0705:29 UT).
One can see that
the electron fluxes shift to even higher energies and
that the low-energy electron fluxes disappear. 

We argue that
the ion-electron decoupling interval
corresponds to the ECL surrounding the X-line. 
The plasma velocities (Fig.~\ref{fig:ET}d) are consistent with
the super-Alfv\'{e}nic electron jets (Fig.~\ref{fig:snapshot}a). 
The profile of the E-T diagram (Fig.~\ref{fig:ET}c) resembles
the two-step profile of the E-S diagram (Fig.~\ref{fig:ET}b).
In particular, one can clearly see
the two-step profile in the second half of the event;
The inner ion-electron decoupling region (before 0705:29 UT) and
the outer region (for example, 0705:29 UT -- 0706:29 UT).
The step-like transition is more evident in the observation,
because we employ an artificially high electron temperature in the PIC simulation.
Immediately after the flow reversal (0702:41 UT),
the satellite observed the highest-energy electron flux,
as indicated by the solid arrow in Figure \ref{fig:ET}c. 
This is consistent with the DR in PIC simulation (Fig.~\ref{fig:ET}b).
In the next few intervals,
the highest-energy fluxes temporarily decreased,
while the low-energy electron fluxes increased.
Probably the electron flux in the pedestal region above the ECL (Fig.~\ref{fig:ET}a) are contaminated,
because the magnetic field was positive ($B_x>0$; not shown) during the interval.
In both PIC simulation and the satellite observation,
we recognize the two-step profile of electron fluxes around the reconnection site.
Both in the ECL and in the ion-electron decoupling interval,
high-energy fluxes are found in the absence of the low-energy fluxes. 
The observation is consistent with our picture of
the high-energy Speiser-accelerated electrons in the ECL and
the low-energy noncrossing electrons outside the ECL.


\section{Discussion and summary}
\label{sec:discussion}

We have investigated
the basic properties of the ECL from various angles:
Fluid quantities, VDFs, trajectories, and compositions.
This allows us to understand the physics in and around the ECL
much more deeply than before.  For example, in the electron jet,
we have found that the electrons are in the $\kappa \lesssim 1$ regime.
They are unmagnetized, and follow Speiser orbits.
Our orbit analysis and composition analysis suggest that
the electron VDF consists of the following two Speiser populations.
One is a global-type Speiser motion traveling through the DR,
and
the other is a local-type Speiser motion picked-up by the outflow exhaust.
Since the Speiser electrons gyrate about $B_z$ for a half gyroperiod near the midplane,
the average electron velocity can be faster in $x$ than the {\bf E}$\times${\bf B} velocity,
$V_{ex} \approx V_{e{\perp}x} > w_x$,
when the reconnecting magnetic fields have out-of-plane ${\pm}y$ components.
This results in the violation of the electron ideal condition,
$\vec{E} + \vec{V}_e \times \vec{B} \ne 0$
in the electron jet.\citep{prit01a,kari07,shay07}
In addition, since electrons travel through chaotic or Speiser orbits,
the VDF is no longer gyrotropic, and therefore
the jet is marked by the nongyrotropy measures.\citep{scudder08,aunai13c,swisdak16}
The $z$-bounce motion during the Speiser motion is responsible for
the phase-hole in $v_z$--$z$ (Fig.~\ref{fig:phase432}e),
similar to the DR (Fig.~\ref{fig:phase432}b).\citep{hori08,chen11}
Since the $z$-bounce speed is smaller than
the rotation speed in the $x$--$y$ plane,
one can see an electron pressure anisotropy
with a stronger perpendicular pressure.\citep{le10a}

\citet{hesse08} argued that
the super-Alfv\'{e}nic electron jet speed can be explained by diamagnetic effects. 
They showed that the force balance is similar to the diamagnetic drift and
that the $z$-profile of the off-diagonal component of the pressure tensor
can be fitted by a gyrotropic pressure model. 
During the Speiser motion,
electrons exhibit the meandering motion.
It is plausible to categorize the meandering motion as the diamagnetic drift,
because the diamagnetic drift is not a guiding center drift.
However, it may not be the best way to discuss
a gyrotropic pressure model inside the ECL,\citep{hesse08}
because most electrons are in the unmagnetized regime of $\kappa<1$.
It is more appropriate to say that
``nongyrotropic electrons carry the diamagnetic-type electric current''
in the ECL.

For ions,
similar semicircle or half-ring-type VDFs by Speiser motions
were reported by hybrid simulations,\citep{nakamura98,lot98}
PIC simulations,\citep{zeni13,hietala15,keizo16}
and satellite observations.\citep{hietala15}
In particular, Ref.~\onlinecite{zeni13} discussed
impacts of Speiser VDFs in PIC simulations in detail. 
They attributed a slow ion flow at a sub-Alfv\'{e}nic speed and
the violation of ion ideal condition to Speiser orbits.
In this work, we attribute
both a fast electron flow at a super-Alfv\'{e}nic speed and
the violation of electron ideal condition
to Speiser motion. 
Ion physics and electron physics are similar, but
some apparent effects are opposite:
The ion flow looks slow while the electron flow looks fast. 
Ref.~\onlinecite{zeni13} further argued
that some ions travel through
figure-eight-shaped regular orbits.\citep{chen86}
We support this discovery
by presenting electron regular orbits. 
The V-shaped path (Fig.~\ref{fig:traj_c}f) corresponds to
a narrow ion channel in the phase-space diagrams
(Figs.~6b and 11b in Ref.~\onlinecite{zeni13}).
Therefore,
both Speiser orbits and regular orbits appear
in the nongyrotropic region in magnetic reconnection,
regardless of plasma species.

We have further introduced a new family of electron orbits,
the noncrossing orbits.
They are attributed to the polarization electric field $E_z$.
Particle motions are organized in a conceptual model in Figure \ref{fig:orbit247}b. 
Similar to conventional orbits,
there exist the noncrossing Speiser orbits and
the noncrossing regular orbits.
The noncrossing Speiser orbits can be further classified into
noncrossing global Speiser orbits and
noncrossing local Speiser orbits (Sec.~\ref{sec:noncrossing}).
As seen in VDFs in Figure \ref{fig:VDF2},
the noncrossing electrons are confined in a low-energy part of the VDFs.
They are the majority in number density (Fig.~\ref{fig:dist}a).
One can order-estimate a typical energy of the noncrossing electrons
($\mathcal{E}_{\rm NC}$) in the following way:
Considering that the plasma density is nearly uniform
over the reconnection site ($\sim n_b$; Fig.~\ref{fig:snapshot}f),
we obtain $n_b(T_i+T_e) \approx \frac{1}{2\mu_0}B_{0}^2$
from the pressure balance across the ECL.
We find that ions sustain most of the perpendicular pressure.
Since the ion bounce motion sustains the polarization field,
the electrostatic potential energy should be
a small fraction ($\delta$) of the ion energy, $\delta \lesssim \mathcal{O}(0.1)$.
The noncrossing electron energy satisfies
$n_b \mathcal{E}_{\rm NC} \lesssim \delta n_b T_i \approx \delta\frac{1}{2\mu_0}B_{0}^2$.
This yields $\mathcal{E}_{\rm NC} \lesssim 2.5 \delta~m_ic_{Ai}^2$,
in agreement with $\mathcal{E}_{\rm NC} \lesssim 0.25$
in Figures \ref{fig:phase432}c, \ref{fig:phase432}f, and \ref{fig:VDF2}fa.

\begin{figure*}[t]
\centering
\includegraphics[width={0.9\textwidth},clip]{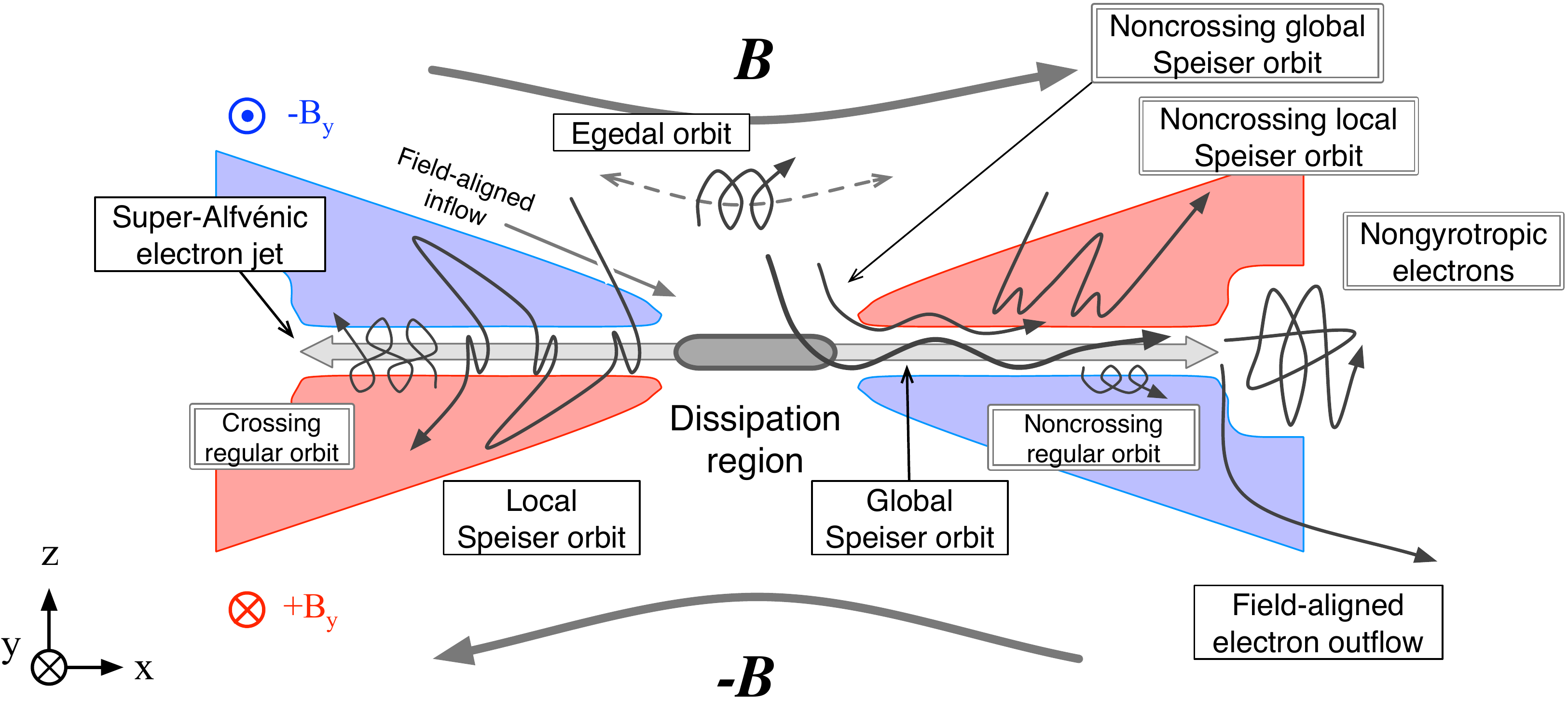}
\caption{(Color online)
A schematic diagram of electron orbits near the reconnection site.
}
\label{fig:diagram}
\end{figure*}

From the macroscopic viewpoint,
the noncrossing electrons challenge
the conventional understanding of
(1) electron mixing and (2) electron heating
during magnetic reconnection.
First, as discussed in Section \ref{sec:comp},
the high-energy electrons from the two inflow regions
mix with each other in the ECL.
On the other hand, the noncrossing electrons do not enter the ECL
due to the Hall field $E_z$.
They start to mix with each other
only downstream of the remagnetization front (Fig.~\ref{fig:dist}b),
where the Hall field $E_z$ disappears (Fig.~\ref{fig:snapshot}c).
In Figure \ref{fig:snapshot}h,
we recognize the enhanced energy dissipation
around the DR,
where the high-energy global Speiser electrons start to mix with each other,
and
near the remagnetization front (indicated by a circle),
where the noncrossing electrons start to mix with each other. 
It is interesting to see that
the nonideal energy transfer corresponds to
these sites of electron mixing.
The relevance between
the dissipation measure $\mathcal{D}_e$ and electron mixing
deserves further research.
Second, the electron heating mechanisms have been
actively studied in the last few years.\citep{phan13b,shay14,haggerty15}
These works reported electrons parallel heating
outside the ECL inside the outflow exhaust.
They implicitly assume the local-type Speiser motion,
while the relevant self-consistent orbits have not been investigated.
In fact, \citet{haggerty15} showed in Figure 2(d) of their paper
that the electron jet region, flanked by the bipolar $E_{\parallel}$ layers,
extends $40 d_i$'s away from the X-line.
This is favorable for the noncrossing electrons.
We expect that
the noncrossing electrons are an integral part of
the electron VDFs of the exhaust region.
The electron heating mechanism should be investigated in more detail,
by considering the noncrossing electrons.

A question is whether
these results are applicable to
magnetic reconnection in the actual magnetotail,
because we have employed artificial parameters in our PIC simulation.
The parameters include
the mass ratio ($m_i/m_e$),
the density ratio ($n_b/n_0$),
the ratio of the plasma and gyro frequencies ($\omega_{pe}/\Omega_{ce}$),
and
plasma beta ($\beta$) in the inflow region.
The mass ratio controls
the relative size of the ECL
within the reconnection site. 
However, as long as the ECL is well resolved,
there is no reason to alter the electron physics. 
We expect that our results scale to the real mass ratio. 
Since tenuous inflow plasma occupies the reconnection site,
the density ratio ($n_b/n_0$) should control
only the build-up time of the ECL structure. 
The frequency ratio $(\omega_{pe}/\Omega_{ce})$
in the inflow region will be important,
because it controls electrostatic properties around the ECL.
\citep{li08,zeni11d,chen12,jara14} 
We expect that
the Hall field $|E_z| \sim c_{Ae,in}B_0 = (\omega_{pe,in}/\Omega_{ce})^{-1} cB_0$
will be important for $(\omega_{pe}/\Omega_{ce})\sim \mathcal{O}(1)$. 
As we estimated in Ref.~\onlinecite{zeni11d},
the inflow frequency ratio in our PIC simulation is
\begin{align}
\frac{\omega_{pe,in}}{\Omega_{ce}} = \frac{\omega_{pe}}{\Omega_{ce}} \Big(\frac{n_b}{n_0}\Big)^{1/2} \approx 1.8.
\end{align}
In the tail lobe,
we expect $\omega_{pe,in}/\Omega_{ce} = 1.6$--$2.3$
for $B=20$ nT and $n_{b}=0.01$-$0.02~{\rm cm}^{-3}$.
Thus, our results will be applicable to the magnetotail reconnection.
The plasma beta $\beta$, in particular,
the electron beta $\beta_e\equiv 2\mu_0 p_e/B^2$, may be another important factor.
In the cold limit of $\beta_e\rightarrow 0$, as in the magnetotail,
we expect that fine structures will be more evident in electron VDFs.\citep{bessho14,shuster15}

The noncrossing electrons can be more pronounced
in magnetic reconnection in different configurations.
We address two favorable cases of
driven-type reconnection and asymmetric reconnection.
In the so-called driven systems,
inflow plasmas are continuously injected toward the reconnection site.
In such a case,
the ions inter-penetrate
deeper than in the undriven case
and therefore
the polarization electric field becomes stronger.
Then we expect more noncrossing electrons on both sides of the midplane.
In fact, \citet{hori08} presented
an electron phase-space diagram across the X-line
in a PIC simulation of driven reconnection (Fig.~5 in Ref.~\onlinecite{hori08}).
The diagram shows clear signatures of noncrossing electrons,
two high-density electron regions in red
inside the ion meandering region. 
In asymmetric systems,
there is often a density gradient across the two inflow regions.
Previous PIC simulations\citep{prit08,tanaka08}
reported a strong normal electric field
on the low-density side of the boundary layer.
This is a polarization field and
this electric field layer overlaps the DR.
In such a case, the low-density side of the DR is
very favorable for noncrossing electrons. 
At the Earth's magnetopause,
magnetic reconnection takes place between
a high-density magnetosheath plasma and
a tenuous magnetospheric plasma,
continuously driven by the solar wind.
We expect that
noncrossing electrons will be a key player for understanding
the physics of magnetopause reconnection.
Since the sunward polarization electric field will be enhanced,
magnetospheric electrons will rarely
mix with magnetospheric plasma in the exhaust region. 

In addition,
the reconnection system may involve
an out-of-plane magnetic field (the so-called guide-field).
In fact, reconnection events with a guide-field have been observed
even in the Earth's magnetotail,\citep{eastwood10}
where we expect antiparallel magnetic fields.
Since a small guide-field alters the ECL structure,\citep{prit04,swisdak05,le13}
the noncrossing electron orbits will be modified accordingly.
A guide-field tends to magnetize electrons even in the DR,
while the polarization electric field persists in the guide-field case. 
Electron orbits in guide-field reconnection
deserve independent research.
We have also ignored the variation in the $y$ direction.
At this point, it is not clear
whether our picture persists in 3D configurations or not.
Nevertheless, it is encouraging that
the satellite observation is consistent with
the simulation with noncrossing electrons
(Sec.~\ref{sec:ET}).


In this work,
we have investigated particle dynamics
in the electron current layer (ECL)
in collisionless magnetic reconnection
in an antiparallel magnetic field.
By tracking self-consistent trajectories
in the two-dimensional PIC simulation,
we have found new classes of electron orbits.
The electron motion in and around the ECL is
much more complicated than we have expected before.
The new orbits, as well as previously known orbits,
are schematically illustrated in Figure \ref{fig:diagram}.
In the inflow region,
electrons are gyrating and fast-bouncing in the parallel direction,
as extensively studied by \citet{egedal05,egedal08}.
Near the separatrices,
some electrons stream along the field lines toward the X-line.
Once electrons enter the DR,
they undergo the Speiser motions of global type.\citep{speiser65}
The electrons slowly turn around to the outflow directions
while bouncing in $z$.
Others travel through the Speiser motions of local-reflection type.
Inside the ECL, there exists
an figure-eight-shaped (crossing) regular orbit.\citep{chen86,zeni13}
The polarization electric field introduces
noncrossing regular orbits on the jet flank and
noncrossing Speiser orbits.
Similar to the traditional Speiser orbits,
the noncrossing Speiser orbits can be categorized as
the global type and the local reflection type,
although their difference is ambiguous.
Downstream of the remagnetization front,
some Speiser electrons remain around the center
as nongyrotropic electrons,
while others travel near the separatrices
in field-aligned electron outflows.

Considering particle orbits,
we have discussed key properties of the electron jet.
The electrons are traveling through Speiser orbits.
The fast bulk speed, electron nonidealness, anisotropy, and nongyrotropy
are consequences of the electron nongyrotropic motion
in the $\kappa \lesssim 1$ regime.
The noncrossing orbits are consistent with
the electron density profile,
the energy-dependent spatial distribution of electrons, and
the electron mixing sites with nonideal energy transfer.
They correspond to the following observational signatures of the ECL:
(1) The super-Alfv\'{e}nic electron jet will be
populated by high-energy nongyrotropic electrons.
(2) The electron density is lower than in the jet flank region.
(3) The electron energy-time diagram will exhibit the two-step profile.
We hope these predictions will be confirmed
by the MMS spacecraft\citep{burch16}
in the second science phase targeting the magnetotail.

\begin{acknowledgments}
The authors acknowledge I. Shinohara for comments.
This work was supported by Grant-in-Aid for Young Scientists (B) (Grant No. 25871054).
The authors acknowledge facilities at Center for Computational Astrophysics,
National Astronomical Observatory of Japan. 
\end{acknowledgments}


\end{document}